\pdfoutput=1
\documentclass{pasa}%

\let\jnl@style=\rm
\def\ref@jnl#1{{\jnl@style#1}}
\def\aj{\ref@jnl{AJ}}                   
\def\araa{\ref@jnl{ARA\&A}}             
\def\apj{\ref@jnl{ApJ}}                 
\def\icarus{\ref@jnl{Icarus}}                 
\def\apjl{\ref@jnl{ApJ}}                
\def\apjs{\ref@jnl{ApJS}}               
\def\ao{\ref@jnl{Appl.~Opt.}}           
\def\apss{\ref@jnl{Ap\&SS}}             
\def\aap{\ref@jnl{A\&A}}                
\def\aapr{\ref@jnl{A\&A~Rev.}}          
\def\aaps{\ref@jnl{A\&AS}}              
\def\azh{\ref@jnl{AZh}}                 
\def\baas{\ref@jnl{BAAS}}               
\def\jrasc{\ref@jnl{JRASC}}             
\def\memras{\ref@jnl{MmRAS}}            
\def\mnras{\ref@jnl{MNRAS}}             
\def\pra{\ref@jnl{Phys.~Rev.~A}}        
\def\prb{\ref@jnl{Phys.~Rev.~B}}        
\def\prc{\ref@jnl{Phys.~Rev.~C}}        
\def\prd{\ref@jnl{Phys.~Rev.~D}}        
\def\pre{\ref@jnl{Phys.~Rev.~E}}        
\def\prl{\ref@jnl{Phys.~Rev.~Lett.}}    
\def\pasp{\ref@jnl{PASP}}               
\def\pasj{\ref@jnl{PASJ}}               
\def\qjras{\ref@jnl{QJRAS}}             
\def\skytel{\ref@jnl{S\&T}}             
\def\solphys{\ref@jnl{Sol.~Phys.}}      
\def\sovast{\ref@jnl{Soviet~Ast.}}      
\def\ssr{\ref@jnl{Space~Sci.~Rev.}}     
\def\zap{\ref@jnl{ZAp}}                 
\def\nat{\ref@jnl{Nature}}              
\def\iaucirc{\ref@jnl{IAU~Circ.}}       
\def\aplett{\ref@jnl{Astrophys.~Lett.}} 
\def\apspr{\ref@jnl{Astrophys.~Space~Phys.~Res.}}
\def\bain{\ref@jnl{Bull.~Astron.~Inst.~Netherlands}} 
\def\fcp{\ref@jnl{Fund.~Cosmic~Phys.}}  
\def\gca{\ref@jnl{Geochim.~Cosmochim.~Acta}}   
\def\grl{\ref@jnl{Geophys.~Res.~Lett.}} 
\def\jcp{\ref@jnl{J.~Chem.~Phys.}}      
\def\jgr{\ref@jnl{J.~Geophys.~Res.}}    
\def\jqsrt{\ref@jnl{J.~Quant.~Spec.~Radiat.~Transf.}}
\def\memsai{\ref@jnl{Mem.~Soc.~Astron.~Italiana}}
\def\nphysa{\ref@jnl{Nucl.~Phys.~A}}   
\def\physrep{\ref@jnl{Phys.~Rep.}}   
\def\physscr{\ref@jnl{Phys.~Scr}}   
\def\planss{\ref@jnl{Planet.~Space~Sci.}}   
\def\procspie{\ref@jnl{Proc.~SPIE}}   

\newcommand{\cm}{~{\rm g~cm}^{-3} } 
\newcommand{\magB}{\mathbf{B}}

\newcommand{\cul}{\mathbf{J}}
\newcommand{\vel}{\mathbf{v}}
\newcommand{\etacm}{~{\rm cm}^{2} ~{\rm s}^{-1} } 

\title[Magnetic field and disk evolution]{Magnetic field and early evolution of circumstellar disks}
\author[Tsukamoto]{Yusuke Tsukamoto$^1$ \\
\affil{$^1$RIKEN, 2-1 Hirosawa, Wako, Saitama, Japan}}%

\jid{PASA}
\doi{10.1017/pas.\the\year.xxx}
\jyear{\the\year}

\usepackage[authoryear]{natbib}

\begin{document}%
\begin{abstract}
The magnetic field plays a central role in the 
formation and evolution of circumstellar disks.
The magnetic field connects 
the rapidly rotating central region with the outer 
envelope and extracts angular momentum from the central region
during gravitational collapse of the cloud core.
This process is known as magnetic braking.
Both analytical and multidimensional simulations have shown that 
disk formation is strongly suppressed by magnetic braking in moderately 
magnetized cloud cores in the ideal magnetohydrodynamic limit.
On the other hand, recent observations have 
provided growing evidence of a relatively large disk 
several tens of astronomical units in size existing in some Class 0
young stellar objects.
This introduces a serious discrepancy between the theoretical study and
observations.
Various physical mechanisms have been proposed
to solve the problem of catastrophic magnetic braking,
such as misalignment between the magnetic field and the rotation
axis, turbulence, and non-ideal effect.
In this paper, we review the mechanism of magnetic braking,
its effect on disk formation and early evolution, 
and the mechanisms that resolve the magnetic braking problem.
In particular, we emphasize  
the importance of non-ideal effects.
The combination of magnetic diffusion and 
thermal evolution during gravitational collapse
provides a robust formation process for
the circumstellar disk at the very early phase of protostar formation.
The rotation induced by the Hall effect can supply a sufficient
amount of angular momentum for typical circumstellar disks 
around T Tauri stars.
By examining the combination of the suggested mechanisms,
we conclude that the circumstellar disks commonly form in the very 
early phase of protostar formation.
\end{abstract}
\maketitle%
\section{INTRODUCTION }
\label{sec:intro}
Circumstellar disks are formed around protostars 
during the gravitational collapse of molecular cloud core. 
Because the disks are the formation sites of planets,
the formation and evolution processes of the disk essentially determine
the initial conditions for planet formation.
Hence, understanding disk formation and evolution
is crucial for constructing a comprehensive theory for planet formation.
An accurate description of the angular momentum evolution is required
to investigate the disk evolution because the centrifugal force mainly
balances the gravitational force of the central protostar.

Formation of a circumstellar disk around a
very young protostar had been believed to be a
natural consequence of angular momentum conservation in 
the gravitationally collapsing molecular cloud core.
Observations of cloud cores have shown that they have
finite angular momentum
\citep[e.g.,][]{1993ApJ...406..528G,2002ApJ...572..238C}.
Many studies of the cloud core collapse
without a magnetic field have been conducted
\citep{1979ApJ...234..289B,1998ApJ...508L..95B,
1998ApJ...495..821T,2003ApJ...595..913M,
2008A&A...482..371C,
2009A&A...495..201A,2009MNRAS.400...13W,
2010ApJ...724.1006M,2012MNRAS.427.1182S,
2012MNRAS.419..760W,2013MNRAS.428.1321T,2013MNRAS.436.1667T},
and it is now well established that a relatively large disk with 
a size of $r\sim 100$ AU is formed during the early phase of 
protostar formation and fragmentation also 
occurs in the unmagnetized cores.

However, the magnetic field changes this 
simple process of disk formation.
During the gravitational collapse, a toroidal magnetic field is 
created and the magnetic tension decelerates the gas rotation, removing
the angular momentum.
This process is known as magnetic braking.
Its importance in circumstellar disk formation was recognized in
the past decade, although there had been several theoretical 
studies regarding magnetic braking 
\citep{1974Ap&SS..27..167G,1979MNRAS.187..311G,
1979ApJ...230..204M,1980ApJ...237..877M},
focusing mostly on the angular 
momentum evolution of molecular clouds or cores.
Simulations in which the ideal magnetohydrodynamics (MHD) 
approximation is adopted and
the magnetic field is aligned with the rotation vector 
have shown that disk formation is almost completely suppressed 
in moderately magnetized cloud cores by magnetic braking
\citep{2003ApJ...599..363A,2007MNRAS.377...77P,
2008ApJ...681.1356M,2008A&A...477....9H}.

Several mechanisms have been suggested
to reduce the magnetic braking efficiency.
For example,  misalignment between the magnetic field
and the rotation vector and turbulence are suggested as 
mechanisms that weaken magnetic braking in the ideal MHD limit 
\citep{2009A&A...506L..29H,2012A&A...543A.128J,2012ApJ...747...21S,2013MNRAS.432.3320S,2013A&A...554A..17J,2013ApJ...774...82L}.
Non-ideal effects (Ohmic diffusion, the Hall effect, 
and ambipolar diffusion), which arise from the finite conductivity
in the cloud core, also serve as mechanisms that change the
magnetic braking efficiency \citep{2009ApJ...706L..46D,2011MNRAS.413.2767M,
2011ApJ...733...54K,
2011ApJ...738..180L,2013ApJ...763....6T,2015ApJ...801..117T,
2015MNRAS.452..278T,2015arXiv150905630M,2015ApJ...810L..26T}.

In this paper, we review recent progress on 
the influence of the magnetic field on the 
formation and early evolution of the circumstellar disk.
The paper is organized as follows.
We review the observed properties of cloud cores in \S 2
and summarize gravitational collapse of cloud cores in \S 3.
The main part of this paper, \S 4 to \S 6, covers
recent studies of disk formation and early evolution 
in magnetized cloud cores.
In \S 7, we summarize our current understanding of disk formation and 
early evolution, and discuss future perspectives.


\begin{figure*}
\label{rot_core}
\includegraphics[width=20pc]{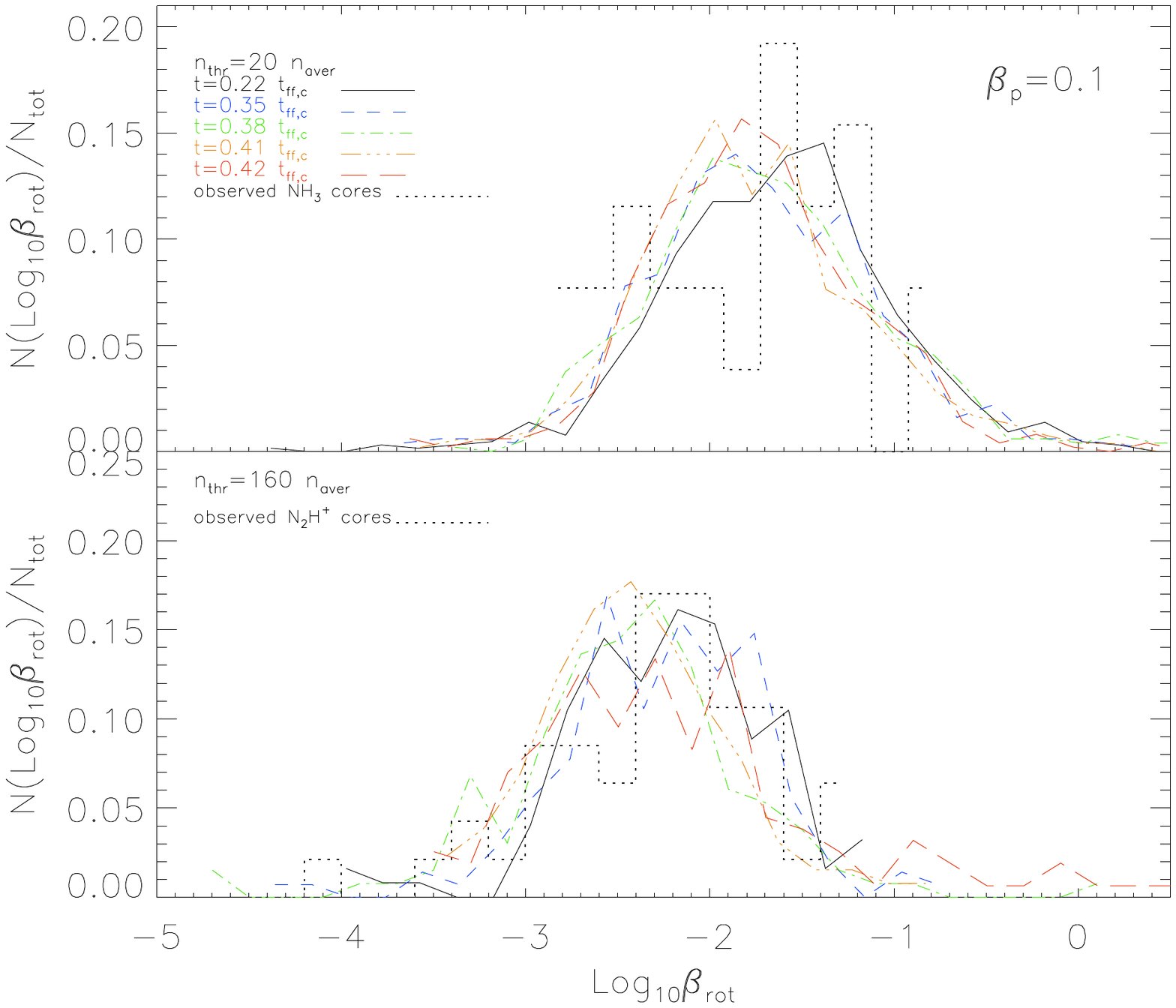}
\includegraphics[width=20pc]{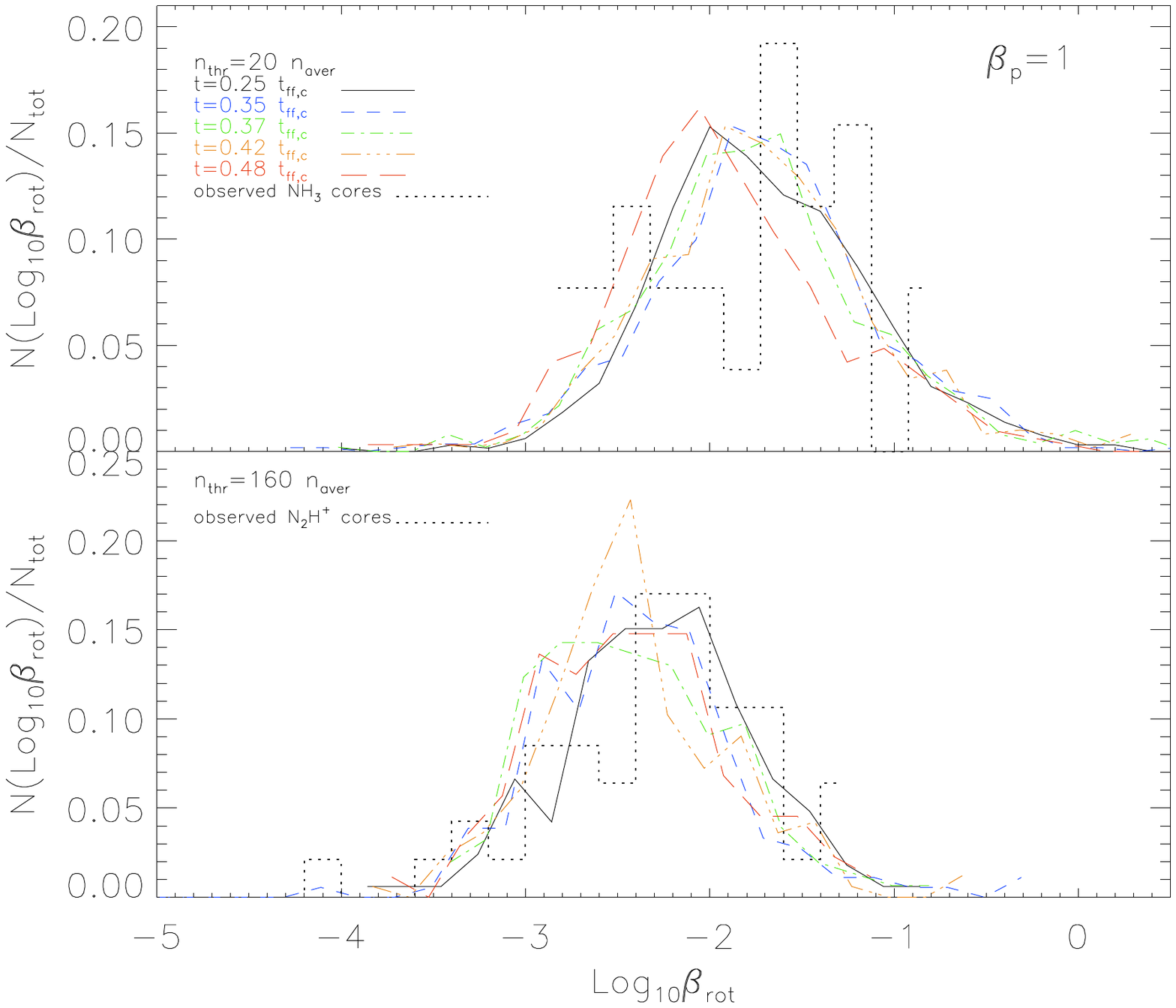}
\caption{
Histogram of $\beta_{\rm rot} (\equiv E_{\rm rot}/E_{\rm grav})$ 
of cloud cores 
obtained using the simulations of \citet{2010ApJ...723..425D} (colored lines)
and observations 
of \citet{1993ApJ...406..528G,1998ApJ...504..207B} and \citet{2002ApJ...572..238C} (black lines).
This figure appears as figure 6 of \citet{2010ApJ...723..425D}.
Colored lines in the upper panels show $\beta_{\rm rot}$ 
with a low density threshold 
for $n_{\rm th}=2.0\times 10^4$ cm$^{-3}$, whereas, 
those in the lower panels show $\beta_{\rm rot}$ 
with a high density threshold for
$n_{\rm th}=8.0\times 10^4$ cm$^{-3}$. The low and high density
thresholds correspond roughly to
the excitation density for the NH$_{3}$ (J-K)=(1,1) transition and 
the N$_{2}$H$^{+}$ (1-0) emission lines, respectively.
The observational results obtained for the
NH$_{3}$ (J-K)=(1,1) transition (upper panels) 
and the  N$_{2}$H$^{+}$ (1-0) emission line (lower panels)
are plotted with black dashed lines.
The NH$_{3}$ core observations are from
\citet{1993ApJ...406..528G} and \citet{1998ApJ...504..207B} 
and the N$_{2}$H$^{+}$ data are from \citet{2002ApJ...572..238C}.
The left and right panels show the results with strong and weak 
initial magnetic fields. The initial plasma $\beta$ in the left and 
right panels are $\beta=0.1$ and  $\beta=1$, respectively.
}
\end{figure*}

\section{Observational properties of molecular cloud cores}
In this section, 
we give an overview of the observational properties of molecular cloud 
cores.

\subsection{Rotation of the cores}
An important parameter of the cloud core is its rotation energy.
Rotation of cloud cores 
is often observationally measured using the velocity gradient 
obtained from the NH$_3$ (1,1) inversion transition line or 
N$_2$H$^+$ (1-0) rotational transition 
line \citep{1993ApJ...406..528G,1998ApJ...504..207B, 
2002ApJ...572..238C,2003A&A...405..639P}.
On the other hand, 
simulations of cloud core formation are performed to theoretically 
investigate core rotation
\citep{2008ApJ...686.1174O,2010ApJ...723..425D}. 
Figure \ref{rot_core} shows the histograms 
of $\beta_{\rm rot}\equiv E_{\rm rot}/E_{\rm grav}$ 
from \citet{2010ApJ...723..425D}, where $E_{\rm rot}$ and $E_{\rm grav}$ are
the rotational and gravitational energy of the core, respectively.
In this figure, both the observation (black dotted lines) 
and simulation results (colored lines) are plotted. The peaks of 
both lines show that the cores typically have a $\beta_{\rm rot}$ 
value of $\sim 0.01$.
Hence, both the observations 
and the simulations suggest that the rotational energy of
a typical cloud core is about 1 \% of its gravitational energy.

\subsection{Turbulence in the cores}
The molecular cloud has a
complex internal velocity structure over a wide range of 
scales that is interpreted as 
turbulent motion \citep{1981MNRAS.194..809L} and,
even at the cloud core scale, 
there exist nonthermal motions \citep{1998ApJ...504..207B}.
\citet{2000ApJ...543..822B} showed that a random Gaussian velocity field
with $ P(k) \propto k^{-4} $ can explain the observed 
rotational properties of the cores.
Note that $ P(k) \propto k^{-4} $ is very similar to the Kolmogorov spectrum
$ P(k) \propto k^{-11/3} $.
Thus, it is expected that turbulence exists in cloud cores
although coherent rotation is often assumed in 
the theoretical study of the cloud core collapse
\citep[e.g.,][]{1998ApJ...508L..95B,2003ApJ...595..913M,
2009MNRAS.400...13W,2011MNRAS.416..591T}.
The turbulent velocity inside the cores is typically subsonic 
\citep{2007prpl.conf...33W}.

\subsection{Magnetic field in the core}
Another important physical quantity is the strength of the magnetic field.
The strength of the magnetic field is often expressed 
using the mass-to-flux ratio relative to the 
critical mass-to-flux ratio \citep{1976ApJ...210..326M}, 
\begin{eqnarray}
\mu=\frac{(M/\Phi)_{\rm core}}{(M/\Phi)_{\rm crit}}=\frac{(M/\Phi)_{\rm core}}{(0.53/3\pi) \sqrt{5/G}}.
\end{eqnarray}
When $\mu<1$, the magnetic pressure is strong enough to
support the cloud core against its self-gravity. 
The critical value, $(M/\Phi)_{\rm crit}=0.53/3\pi \sqrt{5/G}$ is derived for
a spherically symmetric cloud core. This critical value is often used
in theoretical study.
Another critical mass-to-flux ratio is derived for the 
stability of disks and expressed as \citep{1978PASJ...30..671N},
\begin{eqnarray}
\lambda=\frac{(\Sigma/\magB)_{\rm core}}{(\Sigma/\magB)_{\rm crit}}=\frac{(\Sigma/\magB)_{\rm core}}{(4\pi^2 G)^{-1/2}}.
\end{eqnarray}
This is often used in observational study.

\begin{figure}
\begin{center}
\includegraphics[trim=0 0 0 80,width=15pc]{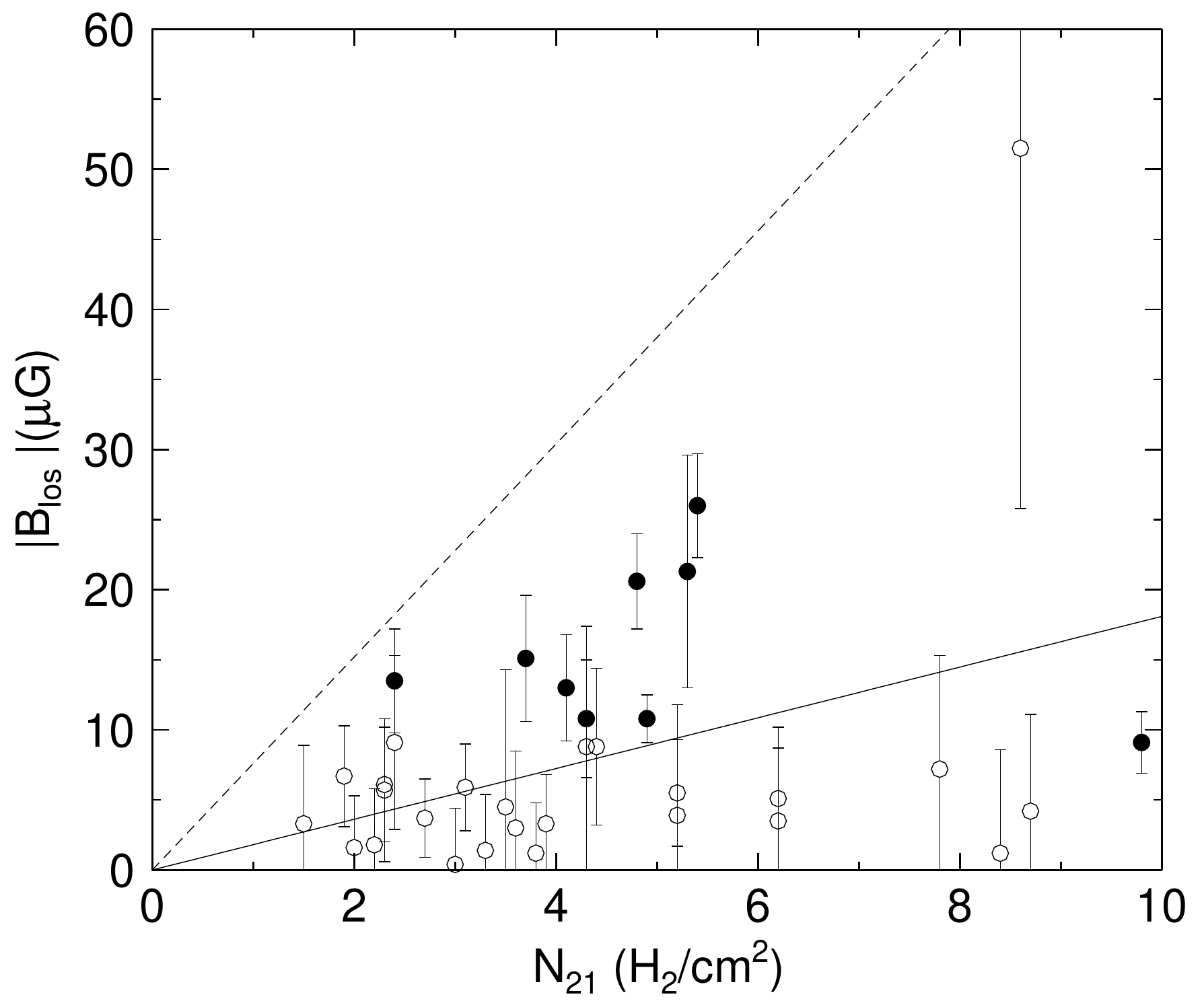}
\caption{
Observed line-of-sight magnetic field strength $B_{\rm los}$
plotted as a function of 
the H$_2$ column density ($N_{21} = 10^{-21} n ({\rm cm^{-2}})$).
This figure appears as  figure 2 of \citet{2008ApJ...680..457T} 
Error bars indicate 1 $\sigma$.
The mass-to-flux ratio normalized by the critical value is given as 
$\lambda=7.6 \times 10^{-21} N_{21}/B_{\rm los}$.
The solid line represents the weighted mean 
value for the mass-to-flux ratio $\lambda=4.8\pm0.4$,
whereas the dashed line represents the value for $\lambda=1$.
}

\label{Bmag_obs}
\end{center}
\end{figure}

The magnetic field strength of the molecular clouds and cores can be
measured using the Zeeman effect \citep{1993ApJ...407..175C,
1996ApJ...456..217C,2008A&A...487..247F,2008ApJ...680..457T,
2012ARA&A..50...29C}. 
Figure \ref{Bmag_obs} shows the 
observation of the magnetic field of the cloud cores 
using the OH Zeeman effect.
This figure appears as figure 2 of \citet{2008ApJ...680..457T}.
They found that the 
mean value of the mass-to-flux ratio of the observed 
cloud cores is $\lambda_{\rm obs}=4.8 \pm 0.4$.
By applying a geometrical correction,
they showed that the mean mass-to-flux ratio of the cloud cores is
$\lambda \sim 2$.
Hence, most cores are 
supercritical, meaning that the magnetic field 
is not strong enough to support the cloud core by  magnetic pressure. 
However, the energy of the magnetic field in cores with $\lambda \sim 2$ 
could be several tens of percent of its
gravitational energy, which is much larger than the rotation velocity.
Therefore, the magnetic field is expected to affect
the gas dynamics during gravitational collapse.

\begin{figure}
\begin{center}
\includegraphics[trim=0 2cm 0 2cm,width=25pc,angle=-270]{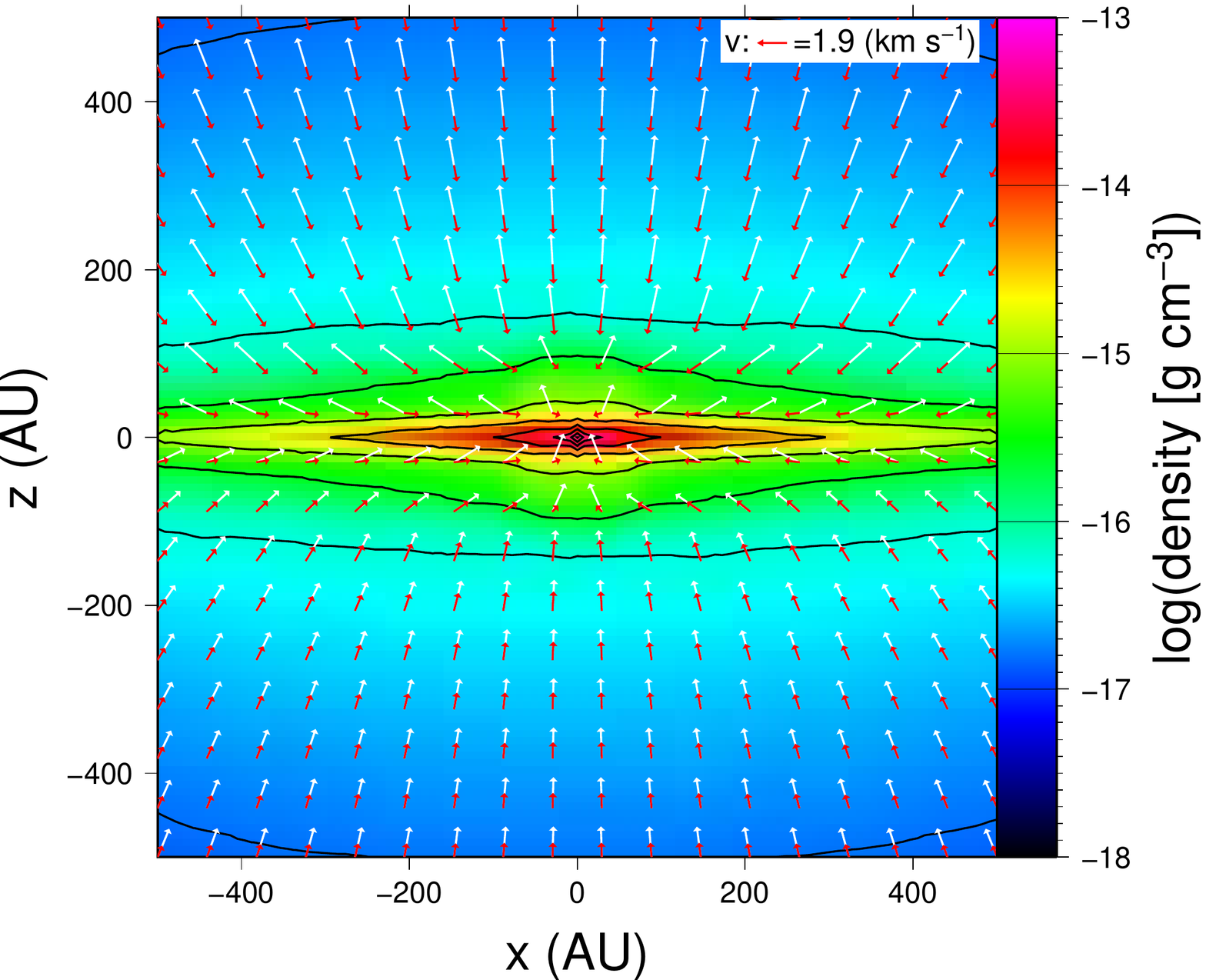}
\caption{
Density structure of the pseudodisk in x-z plane. 
This figure is obtained using simulation results in which all of the
non-ideal effects are considered and 
the magnetic field and rotation vector are parallel.
The simulation corresponds to model Ortho defined 
in \citet{2015ApJ...810L..26T}
and the simulation setup is described in detail in the paper.
At this epoch, the central protostar is formed.
The red and white arrows indicate the velocity field and direction of the
magnetic field, respectively.
}
\label{pseudodisk}
\end{center}
\end{figure}

\section{Gravitational collapse of cloud core}
In this section, 
we discuss gravitational collapse of molecular cloud cores.
Some terminology is also introduced in this section.

Once the core becomes massive enough and gravitationally unstable, 
dynamical collapse of the cloud core begins.
At the beginning of the collapse,
radiation cooling by dust thermal emission is
sufficiently effective,  and the gas temperature
remains almost isothermal at a temperature $T=10$ K.
During this isothermal collapse phase, 
the magnetic field is essentially frozen into the gas.
When the Lorentz force is weak and negligible,
the collapse can be described well as
spherically symmetric collapse.
\citet{1969MNRAS.145..271L} has shown that the
isothermal gravitational collapse proceeds self-similarly.
As a result, the density profile in the isothermal collapse
phase has a central flat profile; the radius is characterized
by the Jeans length $\lambda_{\rm J}$ 
and the outer envelope has $\rho \propto r^{-2} $, as 
shown in \citet{1969MNRAS.145..271L}.
In the spherically symmetric collapse phase, 
the magnetic field evolves as $B\propto \rho^{2/3}$.

As the isothermal spherical collapse proceeds, 
the magnetic field is amplified, and the plasma $\beta\equiv P_{\rm gas}/P_{\rm mag})$ decreases as 
$\beta \propto \rho^{-1/3}$ where $P_{\rm gas}$ and $P_{\rm mag}$ are the 
gas and magnetic pressure, respectively. Hence, at some point,
the Lorentz force becomes effective and 
begins to deflect the gas motion toward the direction parallel to the
magnetic field. This breaks the spherically symmetric collapse.
The gas density increases by the parallel accretion, 
while the magnetic field strength remains almost constant.

As a result of parallel accretion, 
the gas moves to the equatorial plane,
forming a sheet-like structure known as 
a pseudodisk \citep{1993ApJ...417..220G}.
Figure \ref{pseudodisk} shows the density 
map of the pseudodisk formed in the simulation 
of \citet{2015ApJ...810L..26T} for example.
Its radius is typically $r \gtrsim 100$ AU at the protostar
formation epoch.
Because of the inward dragging of the magnetic field,
the magnetic field configuration exhibits an hourglass 
shape \citep{1988ApJ...335..239T,1993ApJ...417..220G} and
a current sheet exists at its midplane.
The hourglass shape of the magnetic field is also inferred
from observations 
\citep{2006Sci...313..812G,2006ApJ...639..965C,2008A&A...490L..39G}.
The figure also shows that the velocity is almost parallel to
the magnetic field except around the midplane indicating that
the gas moves parallel to the magnetic field.
Note that the pseudodisk is not rotationally supported
although its morphology is disk-like.

The gas continues to accrete toward the central region 
mainly through the pseudodisk.
If the disk-like structure is maintained, 
the magnetic field increases as
$B_{\rm c} \propto \rho_{\rm c}^{1/2}$ because 
the central magnetic field and density 
evolve as $B_{\rm c} \propto R^{-2}$ and 
$\rho_{\rm c} \propto R^{-2} H^{-1} \propto R^{-4}$, respectively,
and hence $B_{\rm c} \propto \rho_{\rm c}^{1/2}$.
Here, we assumed that the scale-height of the pseudodisk is given 
by $H_{\rm c}=c_{\rm s}^2/(G \Sigma)=c_{\rm s}/\sqrt{G \rho_{\rm c}}$.

When the central density reaches $ \rho  \sim 10^{-13}{\rm g ~cm^{-3}}$, 
compressional heating
overtakes radiative cooling, and the gas begins to evolve 
adiabatically.
As a result, gravitational collapse temporarily stops and 
a quasi-hydrostatic core, commonly 
known as the first core or the adiabatic core, forms
\citep{1969MNRAS.145..271L, 1999ApJ...510..822M,
2012A&A...543A..60V,2013A&A...557A..90V}. 
In cores with very weak or no magnetic field,
a disk several tens of AU in size can 
form around first core before protostar formation
\citep{1998ApJ...508L..95B,2003ApJ...595..913M,
2009MNRAS.400...13W,2011MNRAS.416..591T,2015MNRAS.446.1175T}.
In the first core phase, the temperature 
evolves as $T \propto \rho^{\gamma-1}$, where $\gamma$ is the 
adiabatic index ($\gamma=5/3$ for $T \lesssim 100$ K, 
and $\gamma=7/5$ for $100 \lesssim T \lesssim 2000$ K).
As we will discuss below, in the first core, magnetic diffusion
becomes effective, and the gas and the magnetic field are temporarily 
decoupled until the central temperature reaches $\sim 1000$ K, and 
thermal ionization provides sufficient ionization.

When the central temperature of the first core 
reaches $\sim2000$ K, the hydrogen molecules begin to dissociate.
This endothermic reaction changes the effective 
adiabatic index to $\gamma_{\rm eff}=1.1$, and
gravitational collapse resumes, which is 
known as the  {\it second collapse}.
Finally, when the molecular hydrogen is completely dissociated, 
the gas evolves adiabatically again, and gravitational collapse 
at the center finishes.
The adiabatic core formed at the center is the protostar (or the second core).
After the protostar forms, it evolves by mass accretion
from the envelope (the remnant of the host cloud core), 
and, at some point, a circumstellar disk is formed around the protostar.

\section{Magnetic braking and suppression of disk formation}
In this section, we review angular momentum transfer 
by the magnetic field, focusing in particular on magnetic braking.
We investigate the most simple case, in which the ideal MHD approximation
is adopted, the magnetic field and rotation axis 
are parallel, and the core rotation is coherent.
The effects of misalignment and turbulence are discussed in \S 5 and
the effects of non-ideal MHD effect are discussed in \S 6.

\subsection{Timescale of magnetic braking}
\label{magneticbraking}
An estimate of the magnetic braking timescale would be 
useful for understanding the basic characteristics of  
magnetic braking.
As shown in many previous studies
\citep{1979ApJ...230..204M,1980ApJ...237..877M,
1985A&A...142...41M,1989MNRAS.241..495N, 1990ApJ...362..202T}, 
the magnetic braking
timescale $t_{\rm b}$ can be estimated as the time in which 
the torsional Alfv\'{e}n waves sweep an amount of gas in the
outer envelope for which the moment of inertia $I_{\rm ext}(t_{\rm b})$ 
equals that of the central region $I_{\rm c}$.
This condition is expressed as
\begin{eqnarray}
I_{\rm ext}(t_{\rm b})=I_{\rm c}.
\end{eqnarray}
By solving this equation for a specified geometry of the central 
region and outer envelope, we can obtain the magnetic braking timescale.

In the simplest geometry, the central collapsing region
is modeled as a uniform cylinder with a density $\rho_{\rm c}$,
radius $R_{\rm c}$, and scale height $H_{\rm c}$ threaded by a uniform
magnetic field parallel to the rotation axis.
The density of the outer envelope,  $\rho_{\rm ext}$ is assumed to be constant.
In this geometry, 
$I_{\rm ext}(t_{\rm b})=\pi \rho_{\rm ext} R_{\rm c}^4 v_{\rm A} t_{\rm b}$ 
and $I_{\rm c}=\pi \rho_{\rm c} R_{\rm c}^4 H_{\rm c}$ 
where $v_{\rm A}$ denotes the Alfv\'{e}n velocity
of the outer envelope.
Thus, $t_{\rm b}$ is given as \citep{1985A&A...142...41M}
\begin{eqnarray}
\label{tb0}
t_{\rm b}=\frac{\rho_{\rm c}}{\rho_{\rm ext}}\frac{H_{\rm c}}{v_{\rm A}}.
\end{eqnarray}
Using the mass of the cylinder, $M=2\pi\rho_{\rm c} R_{\rm c}^2 H_{\rm c}$, and the magnetic 
flux $\Phi=\pi R^2_{\rm c} B$, we can rewrite equation (\ref{tb0}) as
\begin{eqnarray}
\label{tb1}
t_{\rm b} = \left( \frac{\pi}{\rho_{\rm ext}} \right)^{1/2} \frac{M}{\Phi}.
\end{eqnarray}

This shows that the magnetic braking timescale 
in this simple geometry is determined only by 
the mass-to-flux ratio of the central region
and the density of the outer envelope.
This timescale can be regarded as the upper limit 
in the collapsing cloud core because, 
as shown in figure \ref{pseudodisk},
the magnetic field has an hourglass shape
in the gravitationally collapsing cloud core.
In this more realistic configuration, the correction factor ($<1$) 
resulting from the magnetic field geometry 
is multiplied by the braking timescale.

\begin{figure}
\begin{center}
\includegraphics[width=17pc,clip,trim=0cm 1.5cm 0cm 0cm]{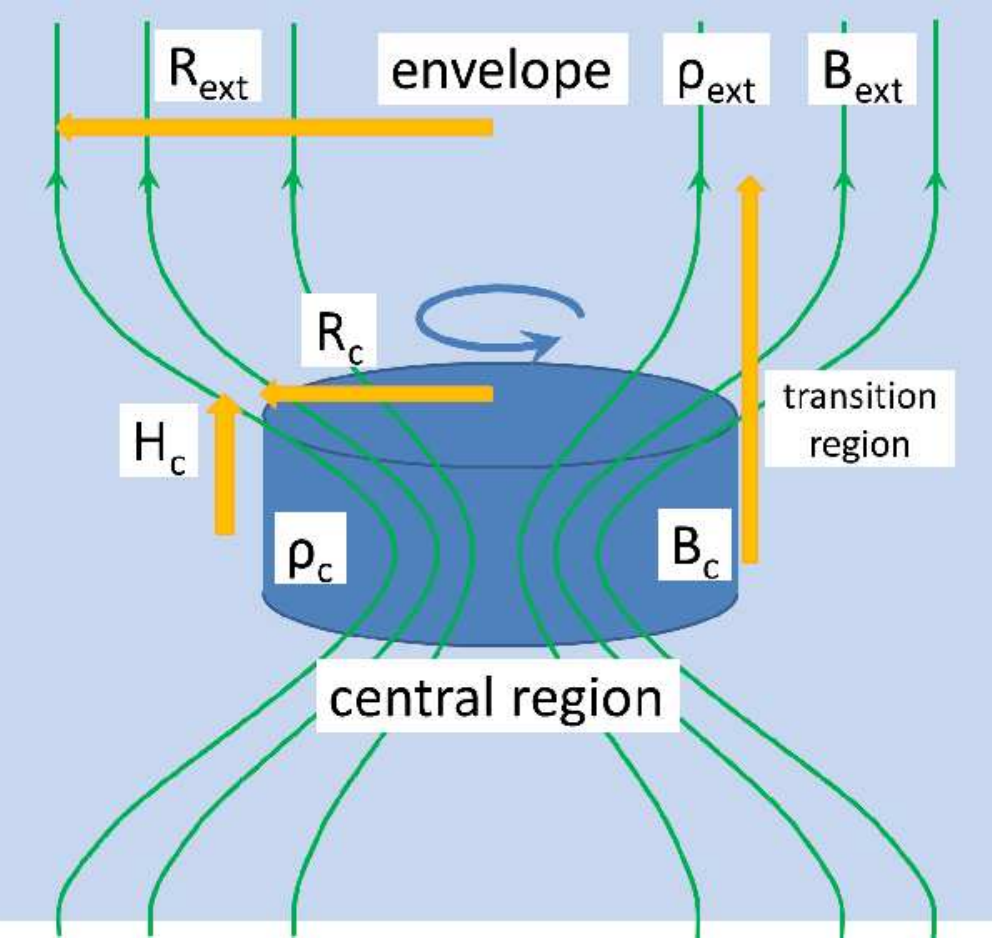}
\caption{
Schematic figure of the geometry assumed in 
the derivation of equations (\ref{eq_ext2}) and (\ref{tb2}). 
$R_{\rm c},~H_{\rm c}$, and $ ~\rho_{\rm c}$ are the radius, 
scale height, and density of the central 
cylinder, respectively. $R_{\rm ext},~\rho_{\rm ext}$
are the radius of flux-tube and density of 
outer envelope, respectively.
}

\label{geometry}
\end{center}
\end{figure}

As illustrated schematically in figure \ref{geometry}, 
in the hourglass configuration,
the magnetic field fans out in the vertical direction.
If we neglect the moment of inertia of the transitional region,
$I_{\rm ext}(t_{\rm b})$ is given as
\begin{eqnarray}
\label{eq_ext2}
I_{\rm ext}(t_{\rm b})=\pi \rho_{\rm ext} R_{\rm ext}^4 v_{\rm A} t_{\rm b}.
\end{eqnarray}
Using $I_{\rm c}=\pi \rho_{\rm c} R_{\rm c}^4 H_{\rm c}$
and equation (\ref{eq_ext2}), we can obtain the 
magnetic braking timescale of the disk
with hourglass magnetic field geometry as \citep{1985A&A...142...41M}
\begin{eqnarray}
\label{tb2}
t_{\rm b,f}=\left( \frac{\pi}{\rho_{\rm ext}} \right)^{1/2} \left(\frac{M}{\Phi}\right)
\left(\frac{R_{\rm c}}{R_{\rm ext}}\right)^2.
\end{eqnarray}
Here, we assume that $R_{\rm ext}=(B_{\rm c}/B_{\rm ext})^{1/2}R_{\rm c}$
because of the conservation of the magnetic flux.
This shows that the magnetic braking timescale could become much shorter
than $t_{\rm b}$ in equation (\ref{tb1}) because $(R_{\rm c}/R_{\rm ext}) < 1$.

The ratio of the radii, $ (R_{\rm c}/R_{\rm ext}) $ is highly uncertain.
Furthermore, the density structure of the envelope evolves with time.
These uncertainties make the analytical treatment of magnetic braking difficult
\citep[see, however,][for example]
{1989MNRAS.241..495N, 1990ApJ...362..202T,2002ApJ...580..987K,
2010A&A...521L..56D,2012A&A...541A..35D}.
Therefore, multidimensional simulation of the collapsing 
cloud core is an important tool 
for investigating the effect of magnetic braking in a realistic 
magnetic field configuration.

\begin{figure}
\begin{center}
\includegraphics[width=20pc]{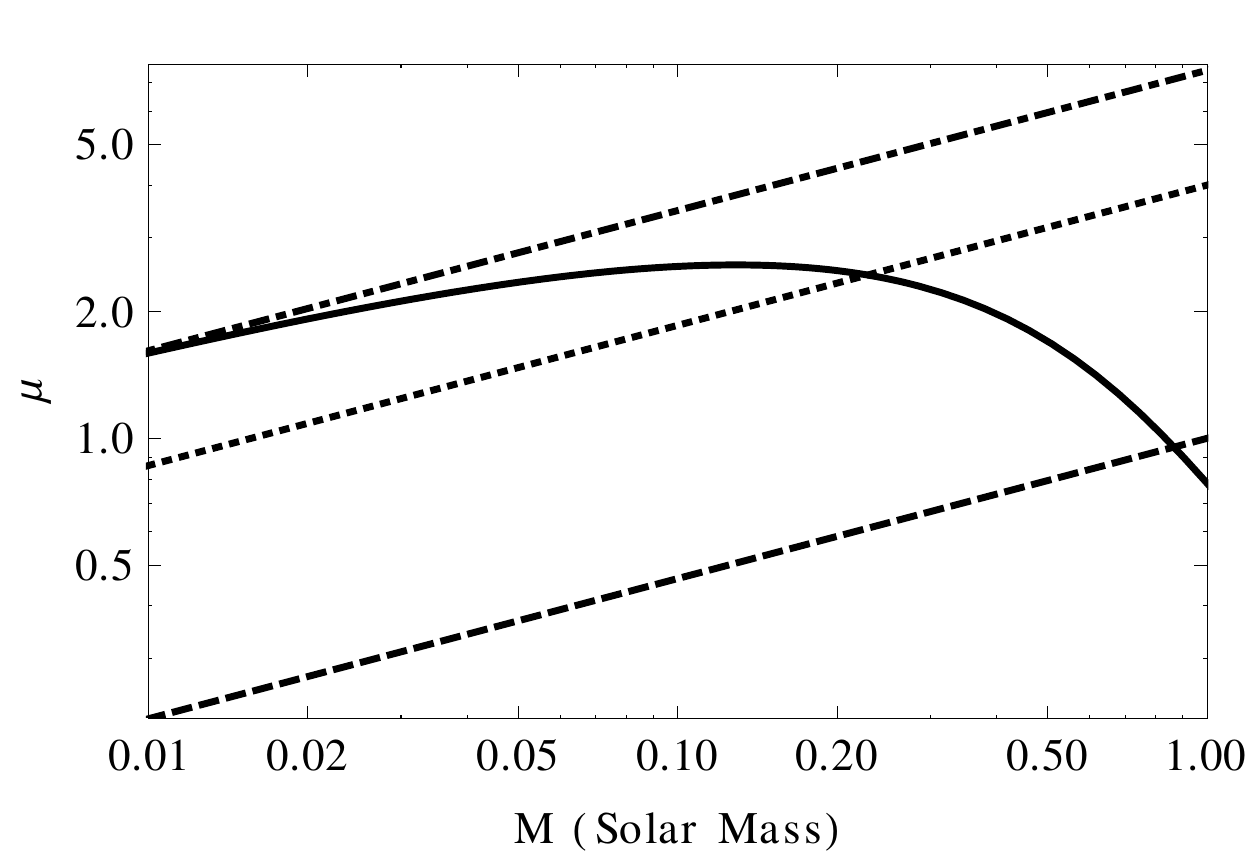}
\caption{
Profile of mass-to-flux ratios of Bonnor-Ebert sphere and uniform sphere 
normalized by the critical 
value $(M/\Phi)_{\rm crit}=0.53/(3\pi) \sqrt{5/G}$ 
as a function of included mass, $M(r)=\int_0^{r} \rho(r') 4 \pi r'^2 dr'$. 
Solid line represents the profile of the Bonnor-Ebert sphere with $\mu=1$ used 
in \citet{2011PASJ...63..555M}.
Dashed, dotted, and dash-dotted lines represent 
the profiles of uniform spheres with $\mu=1~,4~$ and $7.5$, respectively.
Note that \citet{2011PASJ...63..555M} used a different critical value, 
$(M/\Phi)_{\rm crit}=0.48/3\pi \sqrt{5/G}$ 
\citep{1988ApJ...326..208T,1988ApJ...335..239T} and 
the value of the solid line is slightly smaller 
than that shown in figure 2 of the original paper.
}

\label{BE_sphere}
\end{center}
\end{figure}

\subsection{Numerical simulations of magnetic braking}
Using a two-dimensional ideal MHD simulation starting from
cylindrical isothermal cloud cores,
\citet{2000ApJ...528L..41T} clearly showed that much of the angular momentum 
is removed from the central region by magnetic braking and outflow.
He showed that about two-thirds of the initial specific angular 
momentum is removed from the central region during the runaway
collapse phase, and more is removed after the formation of
the first core by outflow and magnetic braking. 
At the end of the simulation, most of the specific angular momentum
has been removed from the central region (a reduction of  $10^4$ from
the initial value).
His simulation clearly indicates the importance of angular momentum
transfer by the magnetic field.

\citet{2003ApJ...599..363A}
showed that magnetic braking in the main accretion
phase is significant and that much of the angular momentum is removed
from the accreting gas 
using  two-dimensional ideal MHD simulations starting from
singular isothermal toroids.
They pointed out that the magnetic braking efficiency
is enhanced by hourglass like magnetic field geometry around the pseudodisk
because the magnetic field is strengthened and
the reduction factor $ (R_{\rm c}/R_{\rm ext})^2 $ in the timescale 
of equation (\ref{tb2}) becomes small.
Because of the two enhancement mechanisms for magnetic braking 
in the pseudodisk, magnetic braking plays an important 
role in the angular momentum evolution of accreting gas.
Note that most of the gas accretes onto the central star through the 
pseudodisk, and angular momentum removal in the pseudodisk strongly
affects formation and evolution 
of the circumstellar disk around the protostar.

This significant removal of angular momentum in the ideal MHD limit 
was later confirmed using two- or three-dimensional simulations
\citep{2006ApJ...641..949B,
2007MNRAS.377...77P,2008A&A...477....9H,2008ApJ...681.1356M,
2011PASJ...63..555M, 2014MNRAS.437...77B}.
These studies focused on the quantitative aspect of magnetic braking, 
{\it i.e.}, how strong a magnetic field is required 
for suppression of the disk formation.
\citet{2007MNRAS.377...77P} showed that
disk formation is strongly suppressed when the mass-to-flux ratio
of entire core is $\mu \lesssim 4$
using three-dimensional smoothed particle hydrodynamics (SPH) 
simulations with a uniform cloud core.
\citet{2008A&A...477....9H} also performed three-dimensional simulations
using a uniform cloud core with adaptive mesh refinement (AMR) code 
RAMSES and also concluded that 
disk formation is suppressed at a slightly greater value 
of the mass-to-flux ratio $\mu \lesssim 5$.
\citet{2008ApJ...681.1356M} 
performed two-dimensional ideal MHD simulations 
using rotating singular isothermal toroids as the initial condition
and showed that circumstellar disk formation is suppressed in
a cloud core with $\mu \lesssim 10$.

Most of studies mentioned above \citep{2000ApJ...528L..41T,
2003ApJ...599..363A,2007MNRAS.374.1347P,
2008A&A...477....9H,2008ApJ...681.1356M,2011PASJ...63..555M} 
used the isothermal or piecewise polytropic 
equation of state (EOS), and 
the influence of the realistic temperature evolution 
on the magnetic braking rate was unclear.
Three-dimensional radiative ideal MHD simulations
with AMR and nested grid codes 
were performed by \citet{2010A&A...510L...3C} and
\citet{2010ApJ...714L..58T}.
They showed that the magnetic braking is significant even when
the radiative transfer is included.
Especially, \citet{2010A&A...510L...3C} showed that the fragmentation 
that occurs in their simulation with $\mu =20$ is suppressed in that with
$\mu = 5$ implying that significant angular momentum removal 
occurs and disk formation is strongly suppressed.
\citet{2014MNRAS.437...77B} conducted radiative ideal MHD simulations of 
a collapsing cloud core using SPH.
They employed a uniform cloud core as the initial condition.
They also showed that disk formation is suppressed when $\mu \lesssim 5$ at
the protostar formation epoch. 
Their results seem to be consistent with previous studies using the 
simplified EOS, and 
radiative transfer would not change
magnetic braking efficiency significantly in the ideal MHD limit.
Note, however, that the fragmentation of the first core or the disk 
is significantly affected by the temperature. Thus, radiative transfer
is important when we consider fragmentation
\citep[][]{2010A&A...510L...3C,2015MNRAS.446.1175T}.

In summary, the previous study indicates
that disk formation is strongly suppressed 
by magnetic braking in the Class 0
phase with an observed magnetic field strength 
$\mu \sim 2$ \citep{2008ApJ...680..457T} in ideal MHD limit
and with aligned magnetic field and rotation vector.
On the other hand, there is growing evidence that a relatively large 
($\sim 50$ AU) disk
exists in some Class 0 young stellar objects \citep{
2013A&A...560A.103M,
2014ApJ...796..131O,
2014Natur.507...78S}
Therefore, obtaining the physical mechanisms that
resolve discrepancy between the
observations and the theoretical study is the main issue
of the recent theoretical study.

\subsection{Consideration of initial conditions}
As we have seen above, the mass-to-flux ratio
of the initial cloud core normalized by the
critical value, $\mu=(M/\Phi)/(M/\Phi)_{\rm crit}$, is often used 
as an indicator of the strength of the magnetic field.
This seems to be reasonable because,
as we have seen in \S \ref{magneticbraking}, the magnetic braking timescale 
is proportional to the mass-to-flux ratio of the {\it central region}.
However, the mass-to-flux ratio $M/\Phi$ is generally a function
of the radius, and the  $M/\Phi$ around the center of 
the initial core can be much larger or smaller than 
the value for the entire cloud core
depending on the density profile and magnetic field profile of 
the initial core.
Thus, we should take care of not only the mass-to-flux ratio of the
entire core but also the initial density  
and initial magnetic field profile of the core 
when we compare the results of previous study.

To illustrate this point,
we show the profiles of the mass-to-flux ratios 
of the two most commonly used initial density profiles, 
i.e., those of a uniform sphere and a Bonnor-Ebert sphere
in figure \ref{BE_sphere}.
The uniform sphere is used in 
\citet{2007Ap&SS.311...75P,2008A&A...477....9H,
2014MNRAS.437...77B,2015ApJ...810L..26T,2015MNRAS.452..278T}
and the Bonnor-Ebert sphere is used mainly in Japanese community,
\citet{2004ApJ...616..266M,2010ApJ...718L..58I,2011MNRAS.413.2767M,2011ApJ...729...42M,2013ApJ...763....6T,2015ApJ...801..117T}.
The profile of the mass-to-flux ratio of the Bonnor-Ebert
sphere depends greatly on its central density, cutoff radius, and total mass.
Thus, the mass-to-flux ratio of the central region is different
among previous studies that used the Bonnor-Ebert sphere.
Here, for example, we select the Bonnor-Ebert sphere 
from model 1 ($\mu=1$) of \citet{2011MNRAS.413.2767M}.

Figure \ref{BE_sphere} shows the profile of mass-to-flux ratio
of Bonnor-Ebert sphere used in \citet{2011PASJ...63..555M} 
and uniform spheres threaded by constant
magnetic field as a function of the included mass, 
$M(r)=\int_0^{r} \rho(r') 4 \pi r'^2 dr'$.
The figure shows that, in the Bonnor-Ebert sphere,
the mass-to-flux ratio around the center of the
core is $\mu(M) \sim 2$ at  $M \sim 0.2 M_\odot$ (solid line) 
even with $\mu=1$ for the entire cloud core.
On the other hand, $\mu(M)$ becomes $ \sim 2$ at $M \sim0.2 M_\odot$ 
in a uniform sphere with $\mu=4$ (dotted line).
If we fix the mass-to-flux ratio at the central region, 
the Bonnor-Ebert sphere with the mass-to-flux ratio of $\mu=1$
corresponds to a uniform sphere with $\mu \sim 7$.
Thus, the mass-to-flux ratios around the center could have
severalfold difference depending on the density profiles.
Note that the magnetic energy is proportional to $|\magB|^2$ and
that the severalfold difference in the 
magnetic field strength results in a difference of more than 
an order of magnitude in the magnetic energy.
Thus, we should pay attention to the initial density profile
when we compare previous results.

An illustrative example regarding this issue can be found in
\citet{2011PASJ...63..555M}.
They conducted three-dimensional 
simulations starting from a supercritical Bonnor-Ebert sphere. 
They showed that, even with a relatively strong magnetic field of 
$\mu=1$, the circumstellar disks can be formed.
This is surprising and seems to contradict other results.
However, it does not contradict to other results.
This difference may come from the difference of 
the magnetic field strength around the center of the cloud core.
In their subsequent paper \citep{2014MNRAS.438.2278M}, 
it is shown that disk formation is more strongly suppressed when 
a uniform sphere is assumed.

\section{Mechanisms that weaken magnetic braking in the ideal MHD limit}
\subsection{Turbulence}
The theoretical study we mentioned above adopted idealized 
cloud cores; {\it i.e.,} 
the core has coherent rotation such as rigid rotation
and the rotation vector and magnetic field are parallel.
A realistic molecular cloud core, however, is expected to have
a turbulent velocity field and
its rotation vector is misaligned from the magnetic field.
In this section, we review the suggested mechanisms that 
weaken the magnetic braking efficiency in the ideal MHD limit.

\citet{2012ApJ...747...21S} suggested that turbulence in the cloud core
weakens magnetic braking. They compared the simulation results for
a coherently rotating core and a turbulent core 
and found that a rotationally supported
disk is formed only in the turbulent cloud core.
Similar results were obtained by \citet{2013MNRAS.432.3320S}.
\citet{2012ApJ...747...21S} pointed out that random motion due to
turbulence causes small-scale magnetic 
reconnections and provides an effective magnetic resistivity that
enables removal of the magnetic flux from the central region.
As a result, in their simulations,
a disk with a size of $r\sim 100$ AU is formed even in ideal MHD limit.

However, their results 
were obtained in the presence of supersonic turbulence
with a Mach number of four, which is much larger than the value expected
from observations (the core typically has subsonic turbulence). 
Furthermore, they employed a uniform grid with a 
relatively large grid size of $\Delta x\sim15$ AU.
In ideal MHD simulations, reconnection occurs at the 
scale of numerical resolution. Thus, a numerical convergence test
is strongly desired to confirm that turbulence-induced reconnection
really plays a role in disk formation.

\citet{2013A&A...554A..17J}
checked the numerical convergence of simulations of turbulent
cloud core collapse with the AMR simulation code RAMSES.
They performed two simulations using exactly the same initial conditions
while varying the numerical resolution 
(they resolved the Jeans length with 10 or 20 meshes) 
and found that the mass of the disk at a given time varies by about a factor 
of two \citep[figure A.1 of][]{2013A&A...554A..17J}.
This result suggests that their simulations do not converge 
and further investigation is desired to quantify the influence of 
turbulent reconnection on disk formation.

\subsection{Misalignment between magnetic field and rotation vector}
Another possible mechanism that weakens the magnetic braking 
is misalignment between the magnetic field and
rotation vector. In many previous studies, it is  assumed 
for simplicity that
the rotation vector is completely aligned with the magnetic field.
However, in real molecular cloud cores, the magnetic field ($\mathbf{B}$) 
and rotation vector ($\mathbf{\Omega}$) would be mutually misaligned.
The recent observations with CARMA suggest that the direction of the
molecular outflows, which may trace the normal direction of the disk,
and the direction of the magnetic field on a
scale of 1000 AU have no correlation \citep{2013ApJ...768..159H}.

In pioneering study on magnetic braking 
\citep{1985A&A...142...41M},
the perpendicular $\mathbf{\Omega} \perp\mathbf{B}$  
configuration was also considered.
The magnetic braking timescale in the the perpendicular configuration 
is given as \citep{1985A&A...142...41M}
\begin{eqnarray}
\label{tb_perp}
t_{\rm b,\perp}=2 \left( \frac{\pi}{\rho_{\rm c}} \right)^\frac{1}{2}\frac{M}{\Phi}.
\end{eqnarray}
In the derivation,
it is assumed that Alfv\'{e}n waves propagate  isotropically
on the equatorial plane, and, as a consequence,
$B(r) \propto r^{-1}$ because of  $\nabla\cdot \magB=0$. 
The ratio of the magnetic braking timescale of parallel and perpendicular
configurations from equation (\ref{tb1}) and (\ref{tb_perp}) is given as
\begin{eqnarray}
\label{tb_ratio1}
\frac{t_{\rm b}}{t_{\rm b,\perp}}=\frac{1}{2}\left(\frac{\rho_{\rm c}}{\rho_{\rm ext}} \right)^{\frac{1}{2}}.
\end{eqnarray}
This shows that the timescale in
perpendicular case is much smaller than that in the
parallel case because $\rho_{\rm c} \gg \rho_{\rm ext}$, meaning that 
the magnetic braking in the
perpendicular case is much stronger than that in the parallel case.
However, in realistic case, fanned-out configuration of magnetic field 
should be considered as shown in figure \ref{geometry}.
Thus, the ratio of the timescale becomes,
\begin{eqnarray}
\label{tb_ratio2}
\frac{t_{\rm b,f}}{t_{\rm b,\perp}}=\frac{1}{2}\left(\frac{\rho_{\rm c}}{\rho_{\rm ext}} \right)^{\frac{1}{2}}\left(\frac{R_{\rm c}}{R_{\rm ext}}\right)^2.
\end{eqnarray}
This shows that magnetic braking timescale of the perpendicular case can be
larger than that of the parallel case when 
$\left(R_{\rm c}/R_{\rm ext}\right)^2 \left(\rho_{\rm c}/\rho_{\rm ext} \right)^{\frac{1}{2}} <1 $.
However, whether 
the magnetic braking in the
perpendicular case is weaker than that in the parallel case  
is not obvious 
because it is difficult to quantitatively compare $(R_{\rm c}/R_{\rm ext})^2$ 
and  $\left(\rho_{\rm c}/\rho_{\rm ext} \right)^{\frac{1}{2}}$ from the 
analytic discussions.

\begin{figure}
\includegraphics[width=20pc]{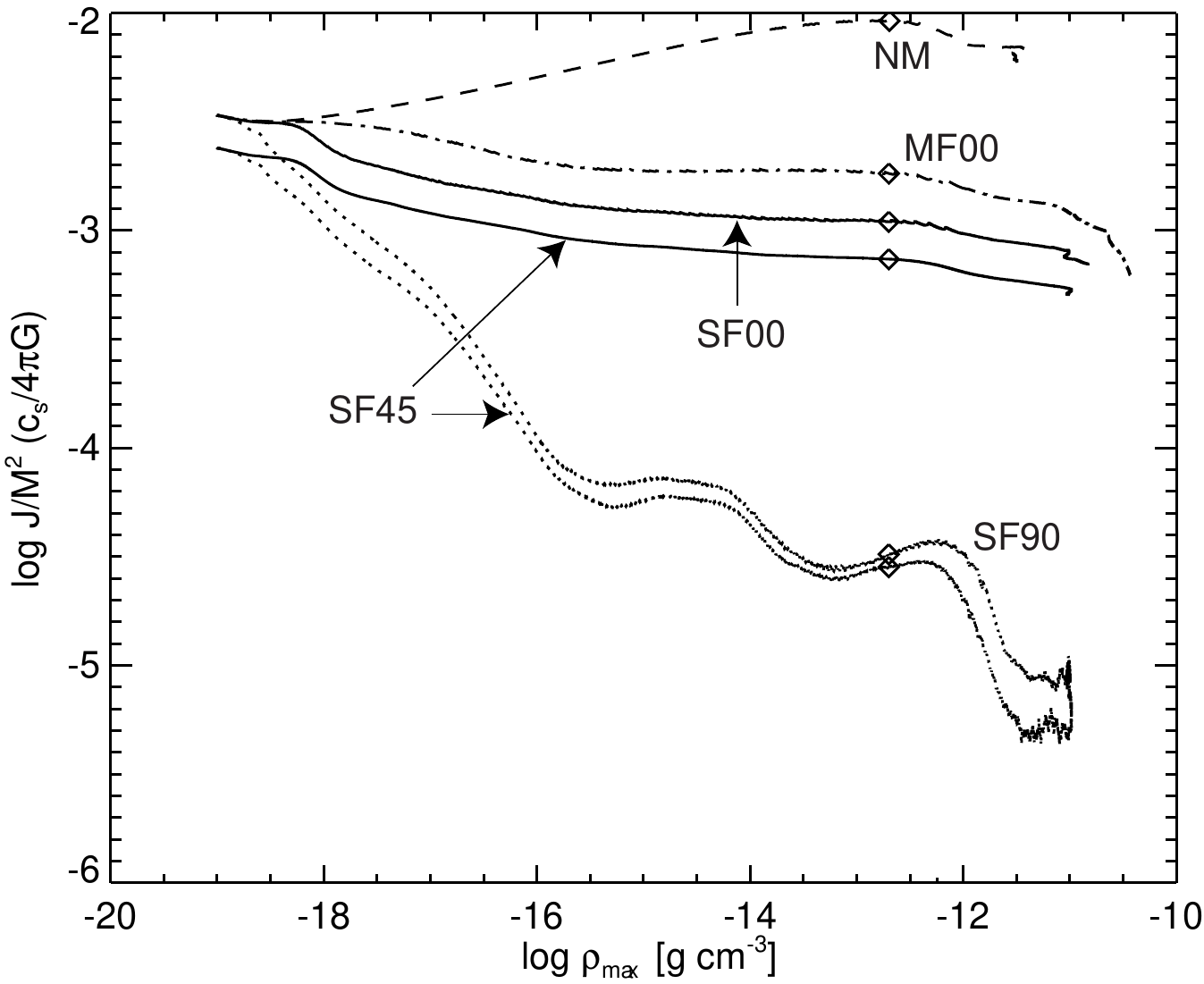}
\caption{
Evolution of central angular momentum  
as a function of maximum (or central) density $\rho_{\rm max}$.
Here, 
$J \equiv \int_{\rho>0.1\rho_{\rm max}}(\mathbf{r \times v}  )\rho~ d \mathbf{V}$
and
$M \equiv \int_{\rho>0.1\rho_{\rm max}}\rho~ d \mathbf{V}$.
This figure appears as figure 12 of \citet{2004ApJ...616..266M}.
Models SF00, SF45, and SF90 denote the simulation results 
with a mutual angle between the initial magnetic field 
and the initial rotation vector of $\theta=0^\circ,~45^\circ$, and $90^\circ$.
The dashed line denotes $J/M^2$ for an unmagnetized simulation.
The solid lines denote the angular momentum parallel 
to the local magnetic field,
$J_\parallel/M^2$, whereas dotted lines denote the
angular momentum perpendicular to the
local magnetic field, $J_\perp/M^2$.
Dash-dotted line denotes $J_\parallel$ for a simulation 
with a weak magnetic field and dashed line denotes $J$ for a simulation
without magnetic field.
Diamonds denote the stage of the first core formation epoch.
Solid line of SF00 and dotted line of SF90 clearly show that
the angular momentum around the central region with a perpendicular
magnetic field is much smaller than that with a parallel magnetic field.
}
\label{J_matsumoto}

\end{figure}

Several multidimensional simulations have been performed to 
investigate the magnetic braking in the  misaligned 
configuration, however, the results are inconsistent
among the previous studies
\citep{2004ApJ...616..266M,2006ApJ...645.1227M,
2009A&A...506L..29H,2012A&A...543A.128J,2013ApJ...774...82L}.
\citet{2004ApJ...616..266M} conducted ideal MHD simulations
of the collapsing cloud core using a Bonnor-Ebert sphere. 
They investigated the angular momentum evolution of the
prestellar collapse phase and reported that the angular 
momentum of the central region is 
more efficiently removed when the magnetic field and rotation vector
are perpendicular. 
This is consistent with the classical estimate of \citep{1979ApJ...230..204M}.
In figure \ref{J_matsumoto}, 
we show the angular momentum evolution of the central
region obtained in \citet{2004ApJ...616..266M}.
The figure shows that the angular momentum in the central region in
the perpendicular case (SF90) is much smaller than that in 
the parallel case (SF00).

\begin{figure*}
\begin{center}
\includegraphics[bb=600 108 1674 684,width=7.5cm]{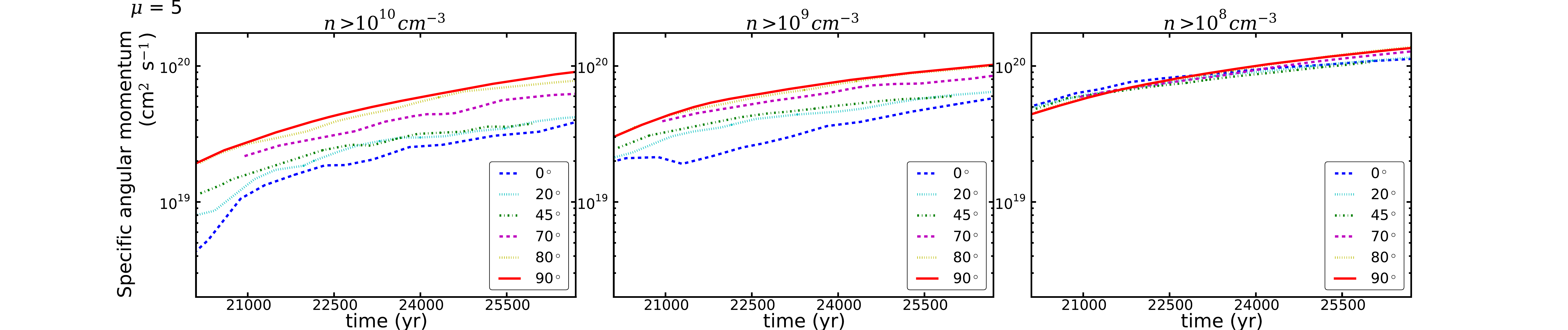}
\vspace{1cm}
\hspace{2cm}
\caption{
Evolution of mean specific angular momentum as a function of time.
This figure appears as figure 4 in \citet{2012A&A...543A.128J}.
Here, the mean specific angular momentum is defined as
$j \equiv \frac{1}{M} \int_{\rho>\rho_{\rm c}}(\mathbf{r \times v}  )\rho~ d \mathbf{V}$
and
$M \equiv \int_{\rho>\rho_{\rm c}}\rho~ d \mathbf{V}$.
Evolution with $\mu=5$ and three different thresholds, 
$\rho_{\rm c}$ that correspond to 
$n=10^{10} {\rm cm^{-3}},~ 10^{9} {\rm cm^{-3}},~ 10^8 {\rm cm^{-3}}$ 
is shown.
}
\label{J_Joos}
\end{center}
\end{figure*}

On the other hand,  
\citet{2009A&A...506L..29H} reported
that the efficiency of the magnetic braking 
decreases as the mutual angle between the magnetic 
field and the rotation axis increases and is minimum 
in the perpendicular configuration
using centrally condensed cloud core with magnetic 
field whose intensity is proportional
to the total column density through the core.
They pointed out that disk formation becomes possible in the
misaligned cloud cores even in the ideal MHD limit.
\citet{2012A&A...543A.128J} also conducted the ideal MHD simulations
with the same density profile of \citet{2009A&A...506L..29H}.
Figure \ref{J_Joos} is taken from figure 4 
of \citet{2012A&A...543A.128J} and
shows that mean specific angular momentum of the central dense region
in a perpendicular core (red lines) is 
about two times larger than that in a parallel core (blue lines).
This is clearly opposite to the result shown in figure \ref{J_matsumoto}.
The influence of misalignment was also 
investigated by \citet{2013ApJ...774...82L} with uniform density sphere.
They also reported that the angular momentum of the central region is
much large in the perpendicular case and
concluded that the disk formation becomes possible when $\mu\gtrsim 4$.
They pointed out that the angular momentum removal by outflow plays
an important role in the parallel configuration.

It is still unclear why the discrepancy 
between the results of 
\citet{2009A&A...506L..29H,2012A&A...543A.128J,2013ApJ...774...82L} 
and \citet{2004ApJ...616..266M} arises.
One possible explanation is the difference in the initial conditions.
As discussed above, the magnetic braking timescale in the 
perpendicular configuration can be larger or smaller than that in the 
parallel configuration depending on the assumptions of the envelope
structure and magnetic field configurations.
Hence, the difference in the initial conditions may explain the discrepancy
although further studies on the effect of misalignment
on the magnetic braking efficiency are required.

\begin{figure}
\begin{center}
\includegraphics[width=20pc]{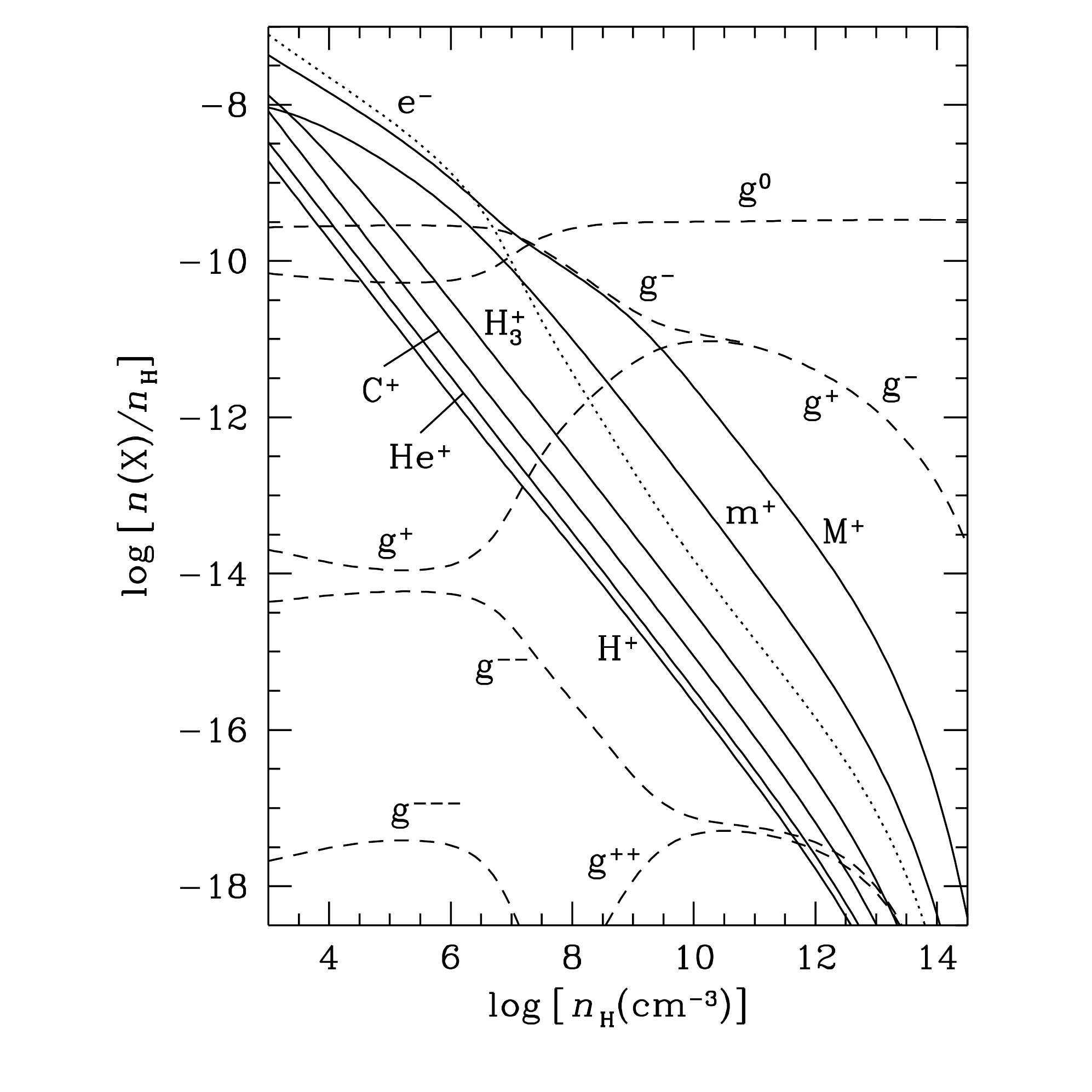}
\caption{
Abundances of various charged particles as a function of the density
of hydrogen nuclei. 
This figure appears as figure 1 of \citet{2002ApJ...573..199N}.
Here, $n_{\rm H}$ denotes the 
number density of hydrogen nuclei. Solid and dotted lines represent
the number densities of ions, and electrons relative to $n_{\rm H}$, respectively.
Dashed lines labeled g$^x$ represent the 
number densities relative to $n_{\rm H}$ of grains of charge $xe$ summed 
over the radius. The ionization rate of a H$_2$ 
molecule by cosmic rays outside the cloud core is taken to be
 $\zeta_0=10^{-17} s^{-1}$. 
M$^+$ and m$^+$ collectively denote 
metal ions such as Mg$^+$, Si$^+$, and Fe$^+$
and molecular ions such as HCO$^+$, respectively.
The MRN dust size distribution \citep{1977ApJ...217..425M} 
with $a_{\rm min}=0.005 \mu m$ and $a_{\rm max}=0.25 \mu m$ is assumed.
}
\label{ionization_degree}
\end{center}
\end{figure}

\begin{figure}
\begin{center}
\includegraphics[width=20pc]{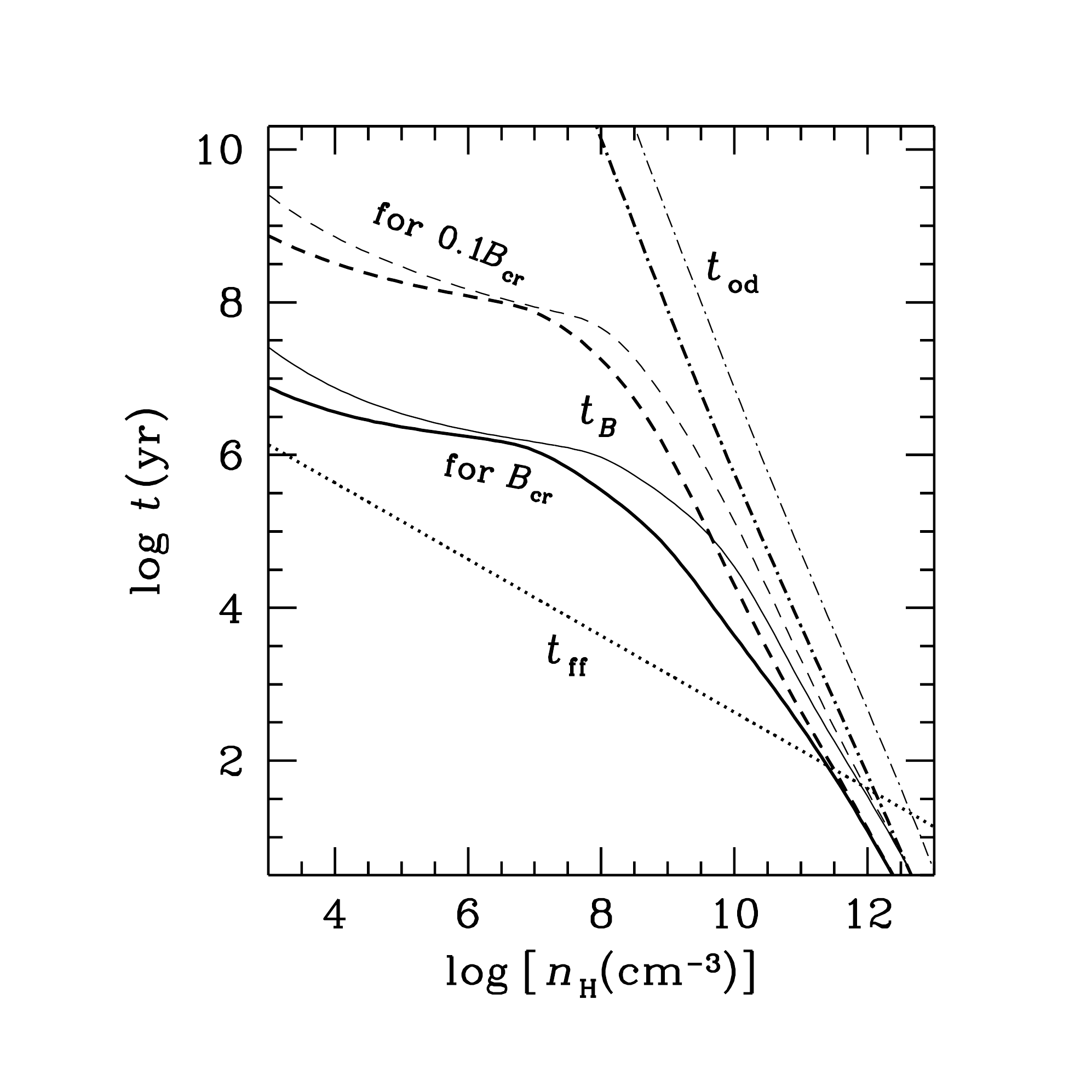}
\caption{
Timescales of magnetic flux-loss for cloud cores.
This figure appears as figure 3 of \citet{2002ApJ...573..199N}.
The flux-loss timescale $t_{\rm B}$ is shown for field strengths of
$B=B_{\rm cr}$ (solid lines) and $B=0.1 B_{\rm cr}$ (dashed lines),
where $B_{\rm cr}$ approximately corresponds 
to the magnetic field strength of $\mu \sim 1$
(the exact value of $B_{\rm cr}$ can be 
found in equation (30) of \citet{2002ApJ...573..199N}).
The Ohmic diffusion time $t_{\rm od}$ is also
shown as dash-dotted lines. 
Two ionization rates by cosmic rays outside the cloud core,  
$\zeta_0=10^{-17} s^{-1}$ (thick lines: standard case) and 
 $\zeta_0=10^{-16} s^{-1}$ (thin lines), are considered. 
The other parameters are the same as in figure \ref{ionization_degree}. 
Dotted line indicates the free-fall time $t_{\rm ff}=(3\pi/(32G\rho))^{1/2}$.
}
\label{tb_tff}
\end{center}
\end{figure}

\section{Influence of non-ideal MHD effects on disk formation} 
So far, we have reviewed the mechanisms that weaken magnetic 
braking in the ideal MHD limit.
In a realistic molecular cloud core, however, the ideal 
MHD approximation, in which infinite conductivity is assumed, 
is not always valid because of the small ionization degree.
Thus, non-ideal effects
may affect the formation and evolution 
of circumstellar disks.
In this section, we review the influence of non-ideal MHD effects on
disk formation.

In a weakly ionized gas, collisions 
between neutral, positively-charged and negatively-charged particles
cause finite conductivity, and non-ideal effects arise.
The non-ideal effects appear as correction terms 
in the induction equation if we neglect the inertia
of the charged particles.
The induction equation with non-ideal terms is given as
\begin{eqnarray}
\label{generalized_induction}
\frac{\partial \magB}{\partial t} &=& \nabla \times(\vel \times \magB)  \\
&-& \nabla \times \left\{ \eta_{\rm O} (\nabla \times \mathbf B) 
+\eta_{\rm H} (\nabla \times \mathbf B)  \times \mathbf {\hat {B}} \right. \\ 
 &-& \left. \eta_{\rm A} ((\nabla \times \mathbf B) \times \mathbf {\hat {B}}) \times \mathbf {\hat {B}}\right\}.
\end{eqnarray}
The second, third, and fourth terms on the right hand side of equation
(\ref{generalized_induction}) describe Ohmic diffusion, the
Hall term, and ambipolar diffusion, respectively.
Here, $\eta_{\rm O}, ~\eta_{\rm H} $, and $ ~\eta_{\rm A} $ are 
the Ohmic, Hall, and ambipolar diffusion coefficients, respectively.
These quantities are calculated from the microscopic force balance
of ions, electrons, and charged dust aggregates.

Detailed calculations of the abundance of charged particles
are required to quantify how the non-ideal effects influence
disk formation.
For example,
we show the evolution of the  abundance of ions, electrons
and charged dusts inside the cloud core as a function of the density
in figure \ref{ionization_degree}. 
This figure appears as figure 1 of \citet{2002ApJ...573..199N}.
The figure shows that the relative abundance of the charged particles
decreases as the density increases.
The figure also shows that the dominant charge carriers are ions and electrons
in the low density region $n_{\rm H} \lesssim 10^6 {\rm cm^{-3}}$ and
g$^+$ and g$^-$ are the dominant carriers
 in the high density region $10^{10}<n_{\rm H} {\rm cm^{-3}}$.

\begin{figure}
\begin{center}
\includegraphics[width=20pc]{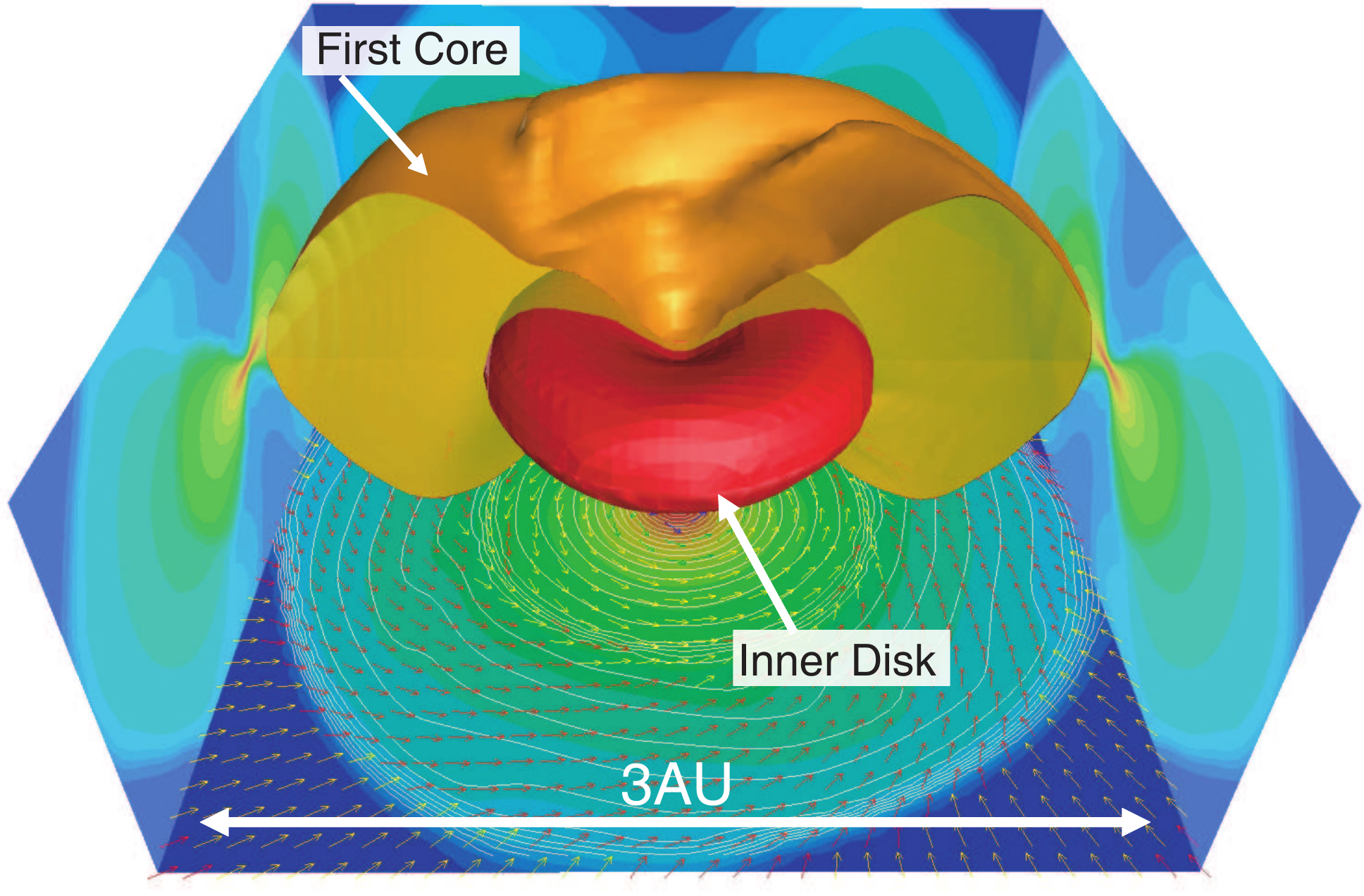}
\caption{
Remnant of the first core (orange isodensity surface) 
and forming circumstellar disk (red isodensity surface) 
plotted in three dimensions. 
This figure appears as figure 3 of \citet{2011MNRAS.413.2767M}.
Density distributions on the x=0, y = 0 and z = 0 planes are
projected onto each wall surface. Velocity vectors on the
z = 0 plane are also projected onto the bottom wall surface.
}

\label{disk_in_FC}
\end{center}
\end{figure}


\begin{figure}
\includegraphics[width=20pc]{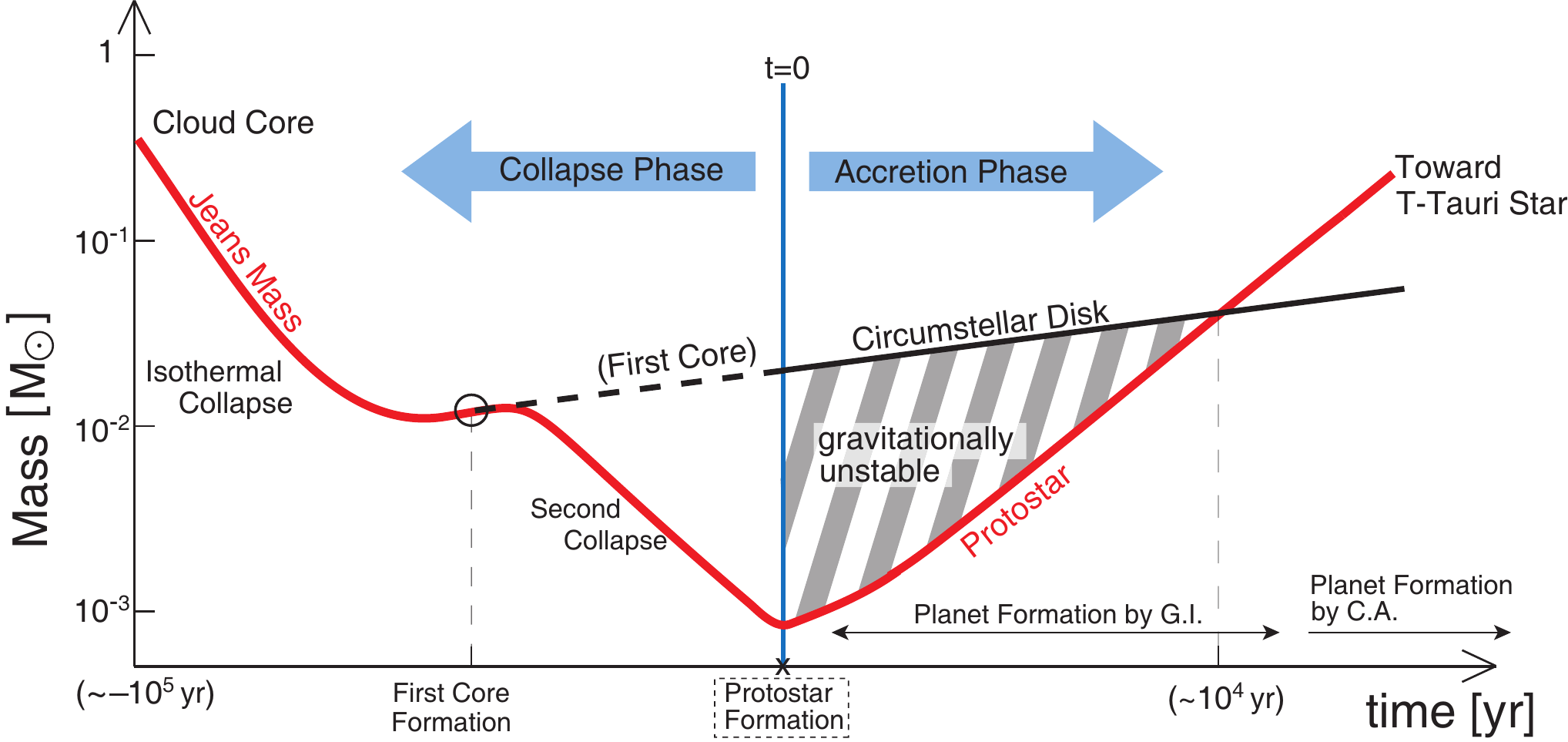}
\caption{
Schematic of evolution of the characteristic mass during
gravitational collapse of the molecular cloud cores.
This figure appears as figure 2 of \citet{2010ApJ...718L..58I}.
The vertical axis denotes mass (in units of solar mass) and 
the horizontal axis denotes time (in years). 
The red curve on the left-hand side indicates the characteristic mass 
of the collapsing molecular cloud core, which 
corresponds to the Jeans mass. 
Note that the mass of the first core is much larger than that 
 of the central protostar at its birth. 
The right-hand side describes the evolution after protostar formation.
Because the first core changes into the circumstellar disk, 
the disk mass remains larger 
than the mass of the protostar in its early evolutionary phase. 
The protostar mass increases monotonically owing to mass accretion from
the disk and becomes larger than the mass of the disk at some point.
}
\label{inutsuka10}
\end{figure}

\subsection{Ohmic and ambipolar diffusion}
\subsubsection{Magnetic flux-loss in the first core phase}
The effect of Ohmic and ambipolar diffusions
in the collapsing cloud core
has been  thoroughly investigated by Nakano and his collaborators 
using an analytic approach \citep{1984FCPh....9..139N,1986MNRAS.221..319N,
1990MNRAS.243..103U,1991ApJ...368..181N,2002ApJ...573..199N}.
They investigated the influence of magnetic diffusion 
during cloud core collapse by
comparing the diffusion timescale of 
magnetic field and the free-fall timescale.

Figure \ref{tb_tff} shows the typical evolution of the
magnetic diffusion timescale in the cloud core. 
The magnetic diffusion timescale becomes
smaller than the free-fall timescale at a density of 
$n_{\rm crit} \sim 10^{11} ~{\rm cm^{-3}}$, and
much of the magnetic flux is removed from the
gas in the central region  when the central density reaches $n_{\rm crit}$.
They pointed out that this flux-loss is caused mainly by Ohmic diffusion.
The critical density varies according to the dust model.
\citet{1991ApJ...368..181N} investigated the dependence 
of the critical density on the dust model and found 
that the critical density  varies in the range of
 $ 10^{10}~{\rm cm^{-3}} \lesssim n_{\rm crit} \lesssim 10^{11} {\rm cm^{-3}}$.

As discussed in \S 3, the pressure-supported 
first core is formed when the central 
density reaches $n \sim 10^{10} ~{\rm cm^{-3}}$, and 
significant flux-loss occurs in the first core phase.
Furthermore, the duration of the first core phase is much longer than
the free-fall timescale and the magnetic flux-loss may occur
at a density less than $n_{\rm crit}$.
Thus, it is expected that the magnetic field and the gas are decoupled in 
the first core and that the magnetic braking is no longer important in it.

\subsubsection{Formation of circumstellar disk in the first core phase}
Multidimensional MHD simulations with magnetic diffusion
have been conducted and have revealed its influence 
on early disk evolution \citep{2009ApJ...706L..46D,2011MNRAS.413.2767M,
2011ApJ...738..180L,2013ApJ...763....6T,2015ApJ...801..117T,
2015MNRAS.452..278T,2015arXiv150905630M}.
As we described above, the magnetic field and the gas
is decoupled in the first core.
Decoupling between the magnetic field and the gas 
in the first core leads to a very important
consequence for disk formation because the 
first core is the precursor of the circumstellar disk.
\citet{2011MNRAS.413.2767M} conducted numerical simulations that
followed formation of the protostar without any sink technique.
They clearly showed that the first core directly becomes 
the circumstellar disk after the second collapse.
In figure \ref{disk_in_FC}, we show the structure of the  
forming circumstellar disk inside the first core 
at the protostar formation epoch.
Because the first core has finite angular momentum and 
magnetic braking is no longer important in it,
the gas cannot accrete directly onto the second core owing to
centrifugal force.
Therefore, the circumstellar disk inevitably
forms just after protostar formation.

was later confirmed by more sophisticated simulations
that included radiative transfer and  Ohmic and ambipolar diffusion
\citep{2013ApJ...763....6T,2015ApJ...801..117T,
2015MNRAS.452..278T,2015arXiv150905630M}.
All of them reported 
formation of a circumstellar disks at the protostar
formation epoch due to magnetic diffusion,
although slight differences exist in the initial size of 
the circumstellar disk 
($1~ {\rm AU}\lesssim r \lesssim 10 ~{\rm AU}$), which may 
arise from differences in the initial conditions, EOS,
or resistivity models.
Because the magnetic field and the gas are inevitably decoupled in
the first core,
we robustly conclude that the circumstellar disk 
with a size of $r \gtrsim 1$ AU is formed at the protostar formation epoch.

The circumstellar disk serves as a reservoir for angular momentum.
As pointed out in the classical theory of an accretion disk
\citep{1974MNRAS.168..603L}, 
the gas accreted onto the disk leaves most of the 
angular momentum in the disk and accretes onto the protostar. 
Therefore, a small disk can grow in the subsequent evolution phase 
even though it is small at its formation epoch.

\subsubsection{Properties and long term evolution of newborn disk}
The newborn circumstellar disk is expected to be more massive than
the newborn protostar at its formation epoch.
This was clearly noted by \citet{2010ApJ...718L..58I}.
Figure \ref{inutsuka10}
shows a schematic figure of the evolution of the 
characteristic mass scale during gravitational collapse and the
accretion phase.
The masses of the newborn protostar and the first core
are roughly determined by the Jeans mass and 
are approximately $10^{-3}~ M_\odot$ and $10^{-2}~ M_\odot$ ,
respectively \citep{1998ApJ...495..346M,1999ApJ...510..822M}.
In addition, the newborn circumstellar disk 
acquires most of the mass of the first core.
Thus, the circumstellar disk is more massive than 
the central protostar at its formation epoch.
In such a massive disk, gravitational instability (GI) serves
as an important angular momentum transfer mechanism.
Later, \citet{2011PASJ...63..555M,2015MNRAS.452..278T} 
confirmed that a newborn disk is actually massive, and
GI may serve as the angular momentum transfer mechanism in the early
phase of circumstellar disk evolution.

The disk evolves by mass accretion from the envelope. 
How the disk size increases in the Class 0 phase 
depends strongly on the amount of angular momentum carried into the disk.
Using long-term simulations with a sink cell,
\citet{2011PASJ...63..555M} showed that a 
disk can grow to the 100 AU scale when the envelope is depleted 
({\it i.e.,} at the end of the Class 0 phase).
Note that magnetic braking becomes weak
once the envelope is depleted because the magnetic braking timescale
depends on the envelope density (see, equations \ref{tb1} and \ref{tb2}).

\begin{figure*}
\begin{center}
\includegraphics[width=40pc]{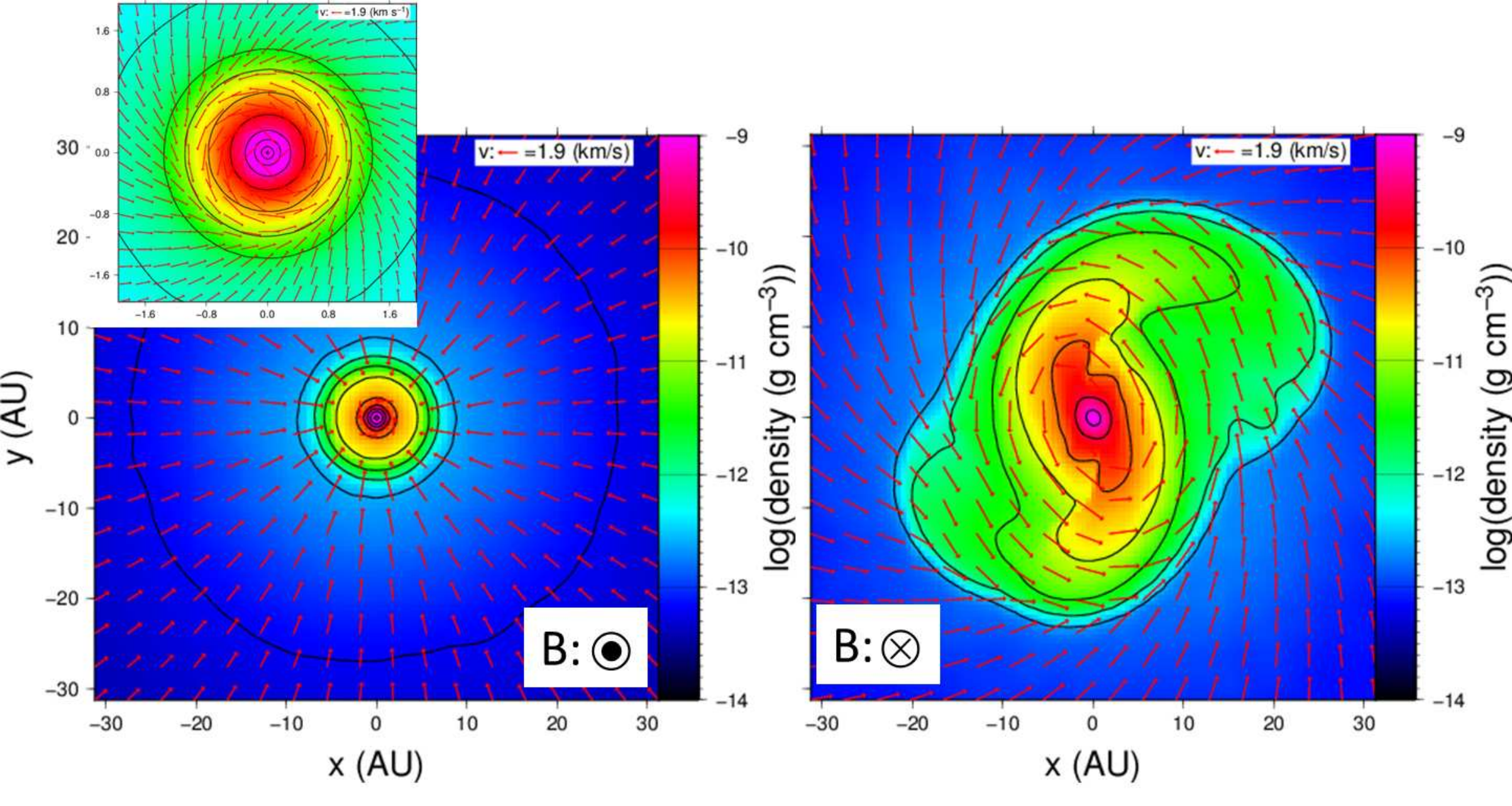}
\caption{
Density map of newborn disks formed in a cloud core with
parallel configuration (left, model Ortho) 
and antiparallel configuration (right, model Para).
This figure is taken from figure 1 of \citet{2015ApJ...810L..26T} but
it has been modified to clarify the formation of the disk in the 
model Ortho.
In the simulations, all the non-ideal effects are considered.
The only difference between the initial conditions of the 
models Ortho and Para is the direction of the magnetic field.
Inset at upper left in the left-hand panel shows an enlarged density map
around the center of the model Ortho. It
shows that a disk
$\sim 1$ AU in size is formed at the center in the parallel case.
The right panel shows that a disk 
$\sim 20$ AU in size is formed at the center in the antiparallel case.
We confirmed that both disks are rotationally supported.
The non-axisymmetric spiral arms in the right panel are created by
gravitational instability. We confirmed that
Toomre's $Q$ value was $Q\sim 1$.
}
\label{hall_disk}
\end{center}
\end{figure*}



\subsection{Hall effect}
The Hall effect has an unique feature  
in that it can actively induce rotation
by generating a toroidal magnetic field from 
a poloidal magnetic field \citep{1999MNRAS.303..239W}.
In this subsection, we review the influence of the Hall effect on
circumstellar disk formation and evolution.

For understanding how magnetic field evolves with the Hall effect,
we rewrite the Hall term in the induction equation as,
\begin{eqnarray}
\label{Hall_rewriten}
-\nabla \times \left\{ \eta_{\rm H} (\nabla \times \mathbf B)  \times \mathbf {\hat {B}}\right\} =\nabla \times \left( \vel_{\rm Hall}  \times \magB \right) ,
\end{eqnarray}
Here, the drift velocity induced by the Hall term is defined as,
\begin{eqnarray}
\label{v_Hall}
\vel_{\rm Hall}=-\eta_{\rm H} \frac{(\nabla \times \magB)_\perp}{|\magB|}=-\eta_{\rm H}\frac{c \cul_\perp}{4 \pi |\magB|},
\end{eqnarray}
where $c$ is the speed of light.
The right-hand side of equation (\ref{Hall_rewriten})
has the same form as the ideal MHD term.
The equations  (\ref{Hall_rewriten}) and (\ref{v_Hall}) show that 
the magnetic field moves along $\cul_\perp$ with a speed of $|\vel_{\rm Hall}|$.

During gravitational collapse, an  
hourglass-shaped magnetic field is generally 
realized (see, figure \ref{pseudodisk}).
In this configuration, a toroidal current exists at the midplane
and the Hall term generates a toroidal magnetic field by twisting the 
magnetic field lines toward the azimuthal direction.
The toroidal magnetic field exerts a toroidal magnetic tension and induces
gas rotation.
Consequently, the gas starts to rotate even when it does not rotate initially.
This phenomenon was actually observed in the simulations 
of \citet{2011ApJ...733...54K} and \citet{2011ApJ...738..180L}.

The characteristic rotation velocity induced by the Hall effect
can be estimated from the Hall drift velocity
$\vel_{\rm Hall}$ because the toroidal component of 
the ideal term and the Hall term
cancel each other out when the rotation velocity is equal to 
the azimuthal component of the Hall drift velocity $v_{\rm Hall,\phi}$.
Thus, the rotation velocity of the gas 
tends to converge to $v_{\phi} = v_{\rm Hall,\phi}$.
Using the numerical simulations in which only Hall effect is considered,
\citet{2011ApJ...733...54K} showed
that the rotation velocity actually converges to $v_{\rm Hall,\phi}$.

Here, we estimate the Hall-induced rotation velocity in the pseudodisk
in which a current sheet exists at the midplane.
The rotation velocity induced by 
the Hall term is roughly estimated as
\begin{eqnarray}
\begin{split}
\label{v_hall1}
v_{\rm \phi} &\sim \frac{\eta_{\rm H}}{|B_{\rm z}|}\frac{|B_{\rm r,s}|}{H} 
\end{split}
\end{eqnarray}
Here, $H$, $B_{\rm z}$ and $B_{\rm r,s}$ are the scale height, 
vertical magnetic field at the midplane, and
radial magnetic field at the surface of the pseudodisk, respectively, and
we assumed $|\nabla \times \magB| \sim |B_{\rm r,s}|/H$.
It is clear from the equation (\ref{v_hall1}) that,
because $\eta_{\rm H}$ is proportional to $|\magB|$, the Hall-induced rotation
velocity is an {\it increasing function} of the strength of the magnetic field.
By employing the monopole approximation $B_{\rm r,s}\sim \Phi_{\rm pdisk}/(2 \pi r^2)$
which is used 
in \citet{1998ApJ...504..247C,2002ApJ...580..987K,
2012MNRAS.427.3188B,2012MNRAS.422..261B}
and using the relation 
of $\Phi_{\rm pdisk}=M_{\rm pdisk}/(\mu_{\rm pdisk} ~(M/\Phi)_{\rm crit})$, 
we can estimate the Hall-induced rotation velocity as,
\begin{eqnarray}
\begin{split}
v_{\rm \phi} &\sim  \frac{\eta_{\rm H}}{|B_{\rm z}|}\frac{|B_{\rm r,s}|}{H} 
= \frac{1}{\mu_{\rm pdisk} (M/\Phi)_{\rm crit}} \frac{\eta_{\rm H}}{|B_{\rm z}|}\frac{M_{\rm pdisk}}{2 \pi r^2 H}\\
&=\frac{\bar{\rho}_{\rm pdisk}}{\mu_{\rm pdisk} (M/\Phi)_{\rm crit}} \frac{\eta_{\rm H}}{|B_{\rm z}|} \\
&\sim 1.0 \times 10^{4} \times \left( \frac{\mu_{\rm pdisk}}{2}\right)^{-1} \left( \frac{\bar{\rho}_{\rm pdisk}}{10^{-14} \cm} \right) \\
& \left( \frac{B_{\rm z}}{10^{-3} ~{\rm G}} \right)^{-1}  \left( \frac{\eta_{\rm H}}{ 10^{18} \etacm }\right) ({\rm cm~s^{-1}}).
\end{split}
\end{eqnarray}
Here, $\Phi_{\rm pdisk},~ \mu_{\rm pdisk},~ M_{\rm pdisk}$, 
and $ \bar{\rho}_{\rm pdisk}$ are the magnetic flux,
the mass-to-flux ratio normalized by the critical value, the mass, and 
the mean density of the pseudodisk, respectively.
Note that $B_{\rm z}$ is the vertical magnetic field at a radius, on the other 
hand, $B_{\rm r,s}$ is determined by the magnetic flux within a radius.
Therefore, we need two different pieces of 
information ($\mu_{\rm pdisk}$ and $B_{\rm z}$) for magnetic field.
Note also that $\Phi_{\rm pdisk}$ (and hence $B_{\rm r,s}$) increases as the total mass in the 
central region is increased by mass accretion
if there is no efficient magnetic flux loss mechanism.
Hence, the Hall-induced rotation would be strengthened in the later
evolution phase.
The corresponding specific angular momentum 
induced by the Hall term is estimated as
\begin{eqnarray}
\begin{split}
\label{specific_j_hall}
j&=r\times v_{\rm \phi}  \\
&\sim 1.5 \times 10^{19} \times  \\
& \left( \frac{r}{100 {\rm AU}} \right) \left( \frac{\mu_{\rm pdisk}}{2}\right)^{-1} \left( \frac{\bar{\rho}_{\rm pdisk}}{10^{-14} \cm} \right) \\
& \left( \frac{B_{\rm z}}{10^{-3} ~{\rm G}} \right)^{-1}  \left( \frac{\eta_{\rm H}}{ 10^{18} \etacm }\right) ({\rm cm^2~s^{-1}}).
\end{split}
\end{eqnarray}

Once a circumstellar disk is formed, 
the accreting gas leaves the most of the angular momentum in
the disk and finally accretes onto the central protostar.
Thus, during protostar formation,
the disk acquires an angular momentum of 
\begin{eqnarray}
\begin{split}
J_{\rm disk, Hall}&=M_*~j \\
&\sim3.1\times10^{52}  \times \\
&\left( \frac{M_*}{M_\odot} \right)\left( \frac{j}{1.5 \times 10^{19}  ~{\rm cm^2~s^{-1}}} \right)  ({\rm g~cm^2~s^{-1}}),
\end{split}
\end{eqnarray}
where $M_*$ is the final mass of the central protostar.

On the other hand,
the total angular momentum of a Keplerian disk 
with $\Sigma\propto r^{-3/2}$ is given as
\begin{eqnarray}
\begin{split}
J_{\rm disk,~Kep}&=\int_{r_{\rm min}}^{R_{\rm disk}} \Sigma(r) r v_{\phi}(r) 2 \pi r dr \\
&\sim \frac{1}{2} M_{\rm disk}\sqrt{G M_* R_{\rm disk}} \\
&\sim 4.4 \times 10^{51} \times \\
&\left( \frac{M_{\rm disk}}{0.01 M_\odot} \right) 
\left( \frac{M_*}{ M_\odot} \right)^{1/2}\left( \frac{R_{\rm disk}}{100 {\rm AU}} \right)^{1/2}  ({\rm g~cm^2~s^{-1}}). 
\end{split}
\end{eqnarray}
Thus, 
the Hall term alone can supply a sufficient amount of
the angular momentum for explaining a circumstellar disk with
a mass and radius of $0.01 M_\odot$ and $100$ AU, respectively,
which roughly correspond to typical values of the disks 
around T Tauri stars \citep{2005ApJ...631.1134A,
2007ApJ...659..705A,2011ARA&A..49...67W}.

In realistic situations, 
the inherent rotation of cloud cores
and magnetic diffusion introduce complicated gas dynamics.
When the rotation of the cloud core is also considered,
a very interesting phenomenon arises.
As we can see from the induction equation, the Hall term
is not invariant against inversion 
of the magnetic field ($\magB \to -\magB$)
and its effect on the gas rotation differs
depending on whether the rotation vector and magnetic 
field of the host cloud core are parallel or antiparallel 
\citep{1999MNRAS.303..239W,2012MNRAS.422..261B,2012MNRAS.427.3188B}.
For $\eta_{\rm H}<0$ which is almost always valid
in the cloud cores,
when the rotation vector and magnetic field are antiparallel,
the  Hall-induced rotation and 
the inherent rotation are in the same direction, and hence,
the Hall term weakens the magnetic braking.
On the other hand, the Hall term strengthens 
the magnetic braking in the parallel case 
because the Hall term induces inverse rotation against the 
inherent rotation of the cloud core.

\citet{2011ApJ...733...54K}
investigated the effect of the Hall term on disk 
formation using two-dimensional simulations.
They focused on the dynamical behavior induced by the Hall term
by neglecting Ohmic and ambipolar diffusion and by employing a constant
Hall coefficient, $Q_{\rm Hall}\equiv \eta_{\rm H} |\magB|$.
They showed that a circumstellar disk 
$r\gtrsim 10$ AU in size can form 
as a result of only the Hall term when the Hall coefficient is
$Q_{\rm Hall} \gtrsim 3\times 10^{20} {\rm cm^2 s^{-1} G^{-1}}$.
Another interesting finding is that the formation of an envelope 
that rotates in the direction opposite to that of disk rotation.
Because of the conservation of the angular momentum, 
the spin-up due to the Hall term at the midplane of the pseudodisk
generates a negative angular momentum flux along the magnetic field line.
This causes spin-down of the upper region,
and the upper region eventually begins 
to rotate in the direction opposite to that of disk rotation.

\citet{2011ApJ...738..180L} 
investigated the effect of the Hall term in two-dimensional simulations
that included all the non-ideal MHD effects 
using a realistic diffusion model and started from uniform cloud cores.
They confirmed that Hall-induced rotation occurs
even when other non-ideal effects are considered. They also showed that
the formation of a counter-rotating envelope.
They showed that the Hall-induced rotation velocity 
can reach $v_\phi\sim 10^5 ({\rm cm ~s^{-1}})$ 
at $r=10^{14} ({\rm cm })$ (this corresponds to the radius of 
the their inner boundary), which means that
the accreting gas has a specific angular momentum of 
 $j \sim 10^{19} ({\rm cm^2 ~s^{-1}})$ 
\citep[figure 11 of][]{2011ApJ...738..180L}.
This is consistent with the value estimated using 
equation (\ref{specific_j_hall}).

\citet{2015ApJ...810L..26T} conducted three-dimensional simulations,
that included all the non-ideal effects as well as radiative transfer.
They followed the first core formation phase and resolved 
protostar formation without any sink technique.
Therefore, their simulations did not suffer from  
numerical artifacts introduced by the sink or inner boundary.
A drawback of this treatment is that they could not follow
the long-term evolution of the disk after protostar 
formation because the numerical timestep became very small.
In figure \ref{hall_disk}, we show a density map of the central regions of
the simulations conducted in \citet{2015ApJ...810L..26T}.
The left panel shows the result of the simulation in which initial magnetic 
field and the rotation vector are in parallel configuration.
On the other hand, the right panel shows that in which initial magnetic 
field and the rotation vector are in antiparallel configuration.
The right panel clearly shows that a disk $\sim 20$ AU in size 
formed at the protostar formation epoch 
On the other hand, the left panel shows that a disk $1$ AU in size
formed even with the parallel configuration.
They also showed that the magnetic field and the gas are decoupled in the 
disk in the right panel, and that 
the magnetic braking is no longer important in it.
Although the disk is formed in both cases, the difference in its size
in the parallel and antiparallel cases
is significant.
Thus, they argued that the disks in Class 0 young stellar objects (YSOs) 
can be subcategorized
according to the parallel and antiparallel nature of their host cloud cores
and suggested that the systems with parallel and antiparallel configurations
should be called as ortho-disks and para-disks, respectively.
They also confirmed that a negatively rotating envelope is formed and
suggested that this envelope may be observable
in future observations of Class 0 YSOs.

Up to the present, \citet{2015arXiv151201597W} 
have made the most comprehensive study regarding the impact of non-ideal 
MHD effects on disk evolution.
They investigated 
the influences of each non-ideal MHD effect both independently
and together using three-dimensional simulations.
They pointed out that, among the three non-ideal effects,
the Hall effect is the most important process for disk size.
This suggests that including the Hall effect is crucial for investigating
the formation and evolution process of circumstellar disks 
in magnetized cloud cores.
They pointed out that an anticorrelation between the size and speed
of the outflow and the size of the disk.
This suggests that angular momentum transfer by the outflow is also
important.
In their simulations, a negatively rotating envelope is also formed.
Note that the negatively rotating envelopes are formed in all multidimensional
simulations with the Hall effect 
\citep{2011ApJ...733...54K,2011ApJ...738..180L, 
2015ApJ...810L..26T,2015arXiv151201597W} and its formation seems to be robust.
Therefore, the detection of a negatively rotating envelope would provide clear 
evidence that the Hall effect actually influences the angular momentum evolution.

\section{Summary and future perspectives}
\subsection{Summary}
In this paper, we reviewed the formation 
and evolution processes of circumstellar disks in magnetized cloud cores,
focusing in particular on the influence of magnetic braking.
In the ideal MHD approximation and with an aligned magnetic field, 
magnetic braking is very efficient and,
circumstellar disk formation is almost completely 
suppressed in a moderately magnetized cloud core  
(its mass-to-flux ratio is $\mu \sim 2$) 
\citep{2003ApJ...599..363A,2008A&A...477....9H,2008ApJ...681.1356M}.
This introduced a serious discrepancy between the observations and theory
and was considered a very serious problem for disk formation theory.

However, various physical mechanisms have been proposed to solve
the problem of catastrophic magnetic braking. 
For example, 
the misalignment between the magnetic field and the rotation
axis \citep{2009A&A...506L..29H,2012A&A...543A.128J} or turbulence 
\citep{2012ApJ...747...21S,2013MNRAS.432.3320S}
in the cloud core may weaken magnetic braking.
Ohmic and ambipolar diffusions remove much of the magnetic flux 
in the first core and 
make it possible for circumstellar disks a few AU in size to form 
at the formation 
epoch of the protostar \citep{2011MNRAS.413.2767M,2013ApJ...763....6T,
2015ApJ...801..117T,2015MNRAS.446.1175T,2015arXiv150905630M}.
The spin-up effect of the Hall term 
increases the specific angular momentum of the accretion flow
and the disk radius at the protostar formation epoch
\citep{2011ApJ...733...54K,2015ApJ...810L..26T}.
A combination of these mechanisms solves the magnetic braking problem,
and we can robustly conclude that the disk is formed in the early evolution
phase of the protostar, 
although its quantitative features, such as the disk radius and mass, 
are still under debate.
Therefore, the simple question of whether a disk can form 
is no longer a central issue,
and we should move on to more specific problems of disk evolution.

\subsection{Future perspectives}
Determining the angular momentum transfer mechanisms in
the Class 0 phase may be the most important unresolved issue.
To data, gravitational instability (GI) and magneto-rotational instability
(MRI) are considered to be the two major mechanisms 
of angular momentum transport within the disk \citep{2011ARA&A..49..195A}.
Further, magneto-centrifugal wind \citep{1982MNRAS.199..883B}
has recently received attention as a mechanism 
that can remove angular momentum from a disk
\citep{2000ApJ...528L..41T,2002ApJ...575..306T,
2013ApJ...769...76B,2013ApJ...772...96B,2015ApJ...801...84G}.
How these mechanisms contribute to disk evolution and
how the relative importance of GI, MRI, and magneto-centrifugal wind
changes during disk evolution are still unclear. 
Surface density and temperature determine whether the disk is 
stable against GI and the size of the MRI dead zone.
Magnetic field strength is closely related to
the saturation level of MRI \citep{2004ApJ...605..321S,2010ApJ...718.1289S}
and strength of magneto-centrifugal wind.
Thus, to answer the question, we must quantitatively investigate 
the long-term evolution of the surface density, temperature, and
magnetic field of disk by considering all the relevant physical mechanisms.

The formation process of binaries or multiples in the Class 0 phase and 
its relation to disk evolution is another important issue.
Fragmentation of the disk or the first core 
is considered a promising mechanism for binary
formation \citep{2003ApJ...595..913M,
2005MNRAS.362..382M,2008ApJ...677..327M,2010ApJ...708.1585K,
2013MNRAS.428.1321T,2015MNRAS.446.1175T}. 
However, for realistic values of the magnetic field and
rotation velocity ($\mu \sim 1$ and $\beta_{\rm rot}\sim 0.01$, respectively), 
the fragmentation in the early phase 
of the protostar formation seems to be strongly suppressed.
In particular, the formation of a binary with a separation of several
tens of AU would be very difficult for realistic values
\citep[see, figure 12 of][]{2008ApJ...677..327M}.
On the other hand, the binary or multiple fraction of solar-type stars 
is about $0.6$, which is quite high \citep{1991A&A...248..485D}.
Furthermore, the median orbital period of binaries is $190$ years 
indicating that the typical separation is several tens of AU.
This introduces a discrepancy between the observations and theoretical studies,
that should be resolved in future studies.
To determine whether fragmentation of the first core or the disk 
is the dominant formation mechanism of multiples,
we should investigate how often the fragmentation of the disk and the
first core can occur in cloud cores
and whether the frequency of fragmentation is sufficient to explain the number
of binaries or multiples fraction.


Dust coagulation in the disk and its impact on disk evolution 
is also an important issue.
It is expected that
dust coagulation occurs and that the size distribution of 
the dust particles 
in the disk changes \citep{2005A&A...434..971D,2009ApJ...698.1122O}.
Once dust coagulation occurs and the dust size distribution changes
during disk evolution, the magnetic resistivity is affected,
and the gas dynamics can also be altered owing to this.
Then, the dust coagulation process may be modified by the gas dynamics.
Therefore, it is possible that the large-scale disk and the
small scale dust distribution co-evolve.
Such a co-evolution process of dust and the disk would be important
not only for the evolution of YSOs but also for formation process
of planetesimals.

\section *{Acknowledgments}
I would like to thank  M. N. Machida, T. Matsumoto, 
S. Okuzumi, S. Inutsuka, K. Iwasaki, 
D. Price, B. Commer{\c c}on, P. Hennebelle, and J. Wurster 
for fruitful discussions.
I also thank the anonymous referee for his/her insightful comments.

\bibliographystyle{apj}
\bibliography{article.bib}

\end{document}